\documentclass[twocolumn]{aastex63}

\usepackage{amsmath}
\usepackage{amsfonts}
\usepackage{graphicx}
\usepackage{rotating}
\usepackage{natbib}
\usepackage{txfonts}
\usepackage{color}
\usepackage{wrapfig}
\usepackage{amssymb}
\usepackage{color}

\shorttitle{Low-metallicity hot core in the LMC} 
\shortauthors{Shimonishi et al.} 

\begin{document}

\title{Chemistry and physics of a low-metallicity hot core in the Large Magellanic Cloud}

\correspondingauthor{Takashi Shimonishi} 
\email{shimonishi@env.sc.niigata-u.ac.jp} 

\author{Takashi Shimonishi} 
\affiliation{Center for Transdisciplinary Research, Niigata University, Ikarashi-ninocho 8050, Nishi-ku, Niigata, 950-2181, Japan}
\affiliation{Environmental Science Program, Department of Science, Faculty of Science, Niigata University, Ikarashi-ninocho 8050, Nishi-ku, Niigata, 950-2181, Japan}

\author{Ankan Das} 
\affiliation{Indian Centre for Space Physics, 43 Chalantika, Garia Station Road, Kolkata 700084, India}

\author{Nami Sakai} 
\affiliation{RIKEN, 2-1 Hirosawa, Wako, Saitama 351-0198, Japan}

\author{Kei E. I. Tanaka} 
\affiliation{Department of Earth and Space Science, Osaka University, Toyonaka, Osaka 560-0043, Japan}
\affiliation{ALMA Project, National Astronomical Observatory of Japan, Mitaka, Tokyo 181-8588, Japan}

\author{Yuri Aikawa} 
\affiliation{Department of Astronomy, Graduate School of Science, The University of Tokyo, 7-3-1 Hongo, Bunkyo-ku, Tokyo 113-0033, Japan}

\author{Takashi Onaka} 
\affiliation{Department of Physics, Faculty of Science and Engineering, 2-1-1 Hodokubo, Hino, Tokyo 191-0042, Japan}
\affiliation{Department of Astronomy, Graduate School of Science, The University of Tokyo, 7-3-1 Hongo, Bunkyo-ku, Tokyo 113-0033, Japan}

\author{Yoshimasa Watanabe} 
\affiliation{College of Engineering, Nihon University, 1 Nakagawara, Tokusada, Tamuramachi, Koriyama, Fukushima 963-8642, Japan}

\author{Yuri Nishimura} 
\affiliation{Institute of Astronomy, The University of Tokyo, 2-21-1, Osawa, Mitaka, Tokyo 181-0015, Japan}
\affiliation{ALMA Project, National Astronomical Observatory of Japan, Mitaka, Tokyo 181-8588, Japan}

\begin{abstract}
We here present the results of 0.1 pc-scale observations in 250 and 350 GHz towards a newly-discovered hot molecular core in a nearby low-metallicity galaxy, the Large Magellanic Cloud (LMC), with the Atacama Large Millimeter/submillimeter Array. 
A variety of C/N/O/Si/S-bearing molecules are detected towards the high-mass young stellar object, ST16. 
A rotating protostellar envelope is for the first time detected outside our Galaxy by SO$_2$ and $^{34}$SO lines. 
An outflow cavity is traced by CCH and CN. 
The isotope abundance of sulfur in the source is estimated to be $^{32}$S/$^{34}$S = 17 and $^{32}$S/$^{33}$S = 53 based on SO, SO$_2$, and CS isotopologues, suggesting that both $^{34}$S and $^{33}$S are overabundant in the LMC. 
Rotation diagram analyses show that the source is associated with hot gas ($>$100 K) traced by high-excitation lines of CH$_3$OH and SO$_2$, as well as warm gas ($\sim$50 K) traced by CH$_3$OH, SO$_2$, $^{34}$SO, OCS, CH$_3$CN lines. 
A comparison of molecular abundances between LMC and Galactic hot cores suggests that organic molecules (e.g., CH$_3$OH, a classical hot core tracer) show a large abundance variation in low metallicity, where the present source is classified into an organic-poor hot core. 
Our astrochemical simulations suggest that different grain temperature during the initial ice-forming stage would contribute to the chemical differentiation. 
In contrast, SO$_2$ shows similar abundances within all the known LMC hot cores and the typical abundance roughly scales with the LMC's metallicity. 
Nitrogen-bearing molecules are generally less abundant in the LMC hot cores, except for NO. 
The present results suggest that chemical compositions of hot cores do not always simply scale with the metallicity. 
\end{abstract} 




\section{Introduction} \label{sec_intro} 
Understanding low-metallicity astrochemistry is crucial to unveil chemical processes in the past universe, where the metallicity was significantly lower than the present-day galaxies. 
Observations of star-forming regions in nearby low-metallicity galaxies and comparative studies of their chemical compositions with Galactic counterparts play an important role for this purpose. 

Hot molecular cores are one of the early stages of star formation and they play a key role in the chemical processing of interstellar molecules, especially for complex molecular species. 
Physically, hot cores are defined as having small source size ($\leq$0.1 pc), high density ($\geq$10$^6$ cm$^{-3}$), and high gas/dust temperature ($\geq$100 K) \citep[e.g.,][]{vanD98,Kur00,vdT04}. 
Characteristic chemistry in hot cores starts from sublimation of ice mantles by stellar radiation and/or shock. 
This leads to the enrichment of gas-phase molecules, and parental species such as methanol (CH$_3$OH) and ammonia (NH$_3$) evolve into larger complex organic molecules (COMs) in warm and dense circumstellar environment \citep[e.g.,][]{NM04,Gar06,Her09,Bal15}. 
Grain surface chemistry also contributes to the formation of COMs upon heating of ice mantles. 
Consequently, hot cores show rich spectral lines in the radio regime. 
Detailed studies of hot core chemistry are thus crucial to understand chemical processes triggered by star-formation activities. 

The Large Magellanic Cloud (LMC) is an excellent target to study interstellar/circumstellar chemistry at low metallicity, thanks to the proximity \citep[49.97 $\pm$ 1.11 kpc,][]{Pie13} and the decreased metallicity environment \citep[$\sim$1/2--1/3 of the solar metallicity; e.g., ][]{Rus92, Duf82, Wes90, And02, Rol02}. 
The low dust-to-gas ratio makes the interstellar radiation field less attenuated, and thus photoprocessing of interstellar medium could be more effective in the LMC than in our Galaxy. 
The environmental differences caused by the deceased metallicity would lead to a different chemical history of star- and planet-forming regions in the LMC and other low-metallicity galaxies. 
Hot core chemistry in the LMC should provide us with useful information to understand chemical complexity in the past metal-poor universe. 

Chemical compositions of interstellar molecules in the LMC have been studied extensively. 
Molecular-cloud-scale chemistry ($\lesssim$10 pc) has been investigated by radio single-dish observations \citep[e.g.,][]{Joh94, Chi97, Hei99, Wan09, Par14, Par16, Nis16, Tan17}. 
Interferometry observations in millimeter have probed distributions of dense molecular gas at a clump scale \citep[a few pc; e.g.,][]{Won06, Sea12, And14}. 
The Atacama Large Millimeter/submillimeter Array (ALMA) has provided us with an unprecedented sensitivity and spatial resolution to study physical properties of dense molecular gas around young stellar objects (YSOs) at a subparsec scale \citep[e.g.,][]{Ind13, Fuk15, Sai17, Nay18}. 
For solid state molecules, compositions of ice mantles have been probed by infrared spectroscopic observations towards embedded YSOs \citep[e.g.,][]{vanL05, ST, ST10, ST16, Oli09, Oli11, Sea11}

Chemistry of hot cores at low metallicity is now emerging with discoveries of extragalactic hot cores in the LMC with ALMA \citep{ST16b, Sew18}. 
The formation of organic molecules in low-metallicity environments is one of the important issues in recent astrochemical studies of low-metallicity star-forming regions. 
\citet{ST16b} reported that organic molecules such as CH$_3$OH, H$_2$CO, and HNCO towards a hot molecular core in the LMC (ST11) is underabundant by 1--3 orders of magnitude compared to Galactic hot cores. 
On the other hand, \citet{Sew18} reported the detection of CH$_3$OH and even larger COMs toward other different hot molecular cores in the LMC (N113 A1 and B3). 
In contrast to the ST11 hot core, they found that the molecular abundances of COMs in the N113 hot cores are scaled by the metallicity of the LMC and comparable to those found at the lower end of the range in Galactic hot cores. 
Because of the limited number of current samples and the limited frequency coverage, chemical processes to form organic molecules in low-metallicity hot cores are still an open question. 
Besides those observational approaches, astrochemical simulations of low-metallicity hot core chemistry are presented in \citet{Bay08,Ach18}. 
Observational efforts to identify and analyze further low-metallicity hot cores are important to constrain various uncertainties involved in astrochemical models. 

In this paper, we report the results of high spatial resolution submillimeter observations with ALMA towards a high-mass YSO in the LMC, and present the discovery of a new hot molecular core. 
Section \ref{sec_tarobsred} describes the details of the target source, observations, and data reduction. 
The obtained molecular line spectra and images are described in Section \ref{sec_res}. 
Derivation of physical quantities and molecular abundances from the present data is described in Section \ref{sec_ana}. 
Properties of the observed hot core and the comparison of the molecular abundances with those of other LMC and Galactic hot cores are discussed in Section \ref{sec_disc}, where astrochemical simulations of low-metallicity hot cores are also presented. 
Distributions of molecules in a rotating protostellar envelope and an outflow cavity, as well as isotope abundances of sulfur in the source, are also discussed in Section \ref{sec_disc}. 
The conclusions are given in Section \ref{sec_sum}.

\begin{deluxetable*}{ l c c c c c c c c c}
\tablecaption{Observation summary \label{tab_Obs}} 
\tablewidth{0pt} 
\tabletypesize{\footnotesize} 
\tablehead{
\colhead{}   & \colhead{Observation} &  \colhead{On-source} & \colhead{Mean}                             & \colhead{Number}   &  \multicolumn{2}{c}{Baseline}     &  \colhead{}                                                  &  \colhead{}                                         &  \colhead{Channel}  \\
\cline{5-6}  
\colhead{}   & \colhead{Date}             &  \colhead{Time}         &  \colhead{PWV\tablenotemark{a}} & \colhead{of}              &  \colhead{Min} & \colhead{Max} & \colhead{Bem size\tablenotemark{b}}        & \colhead{MRS\tablenotemark{c}}     &  \colhead{Spacing}  \\
\colhead{}   & \colhead{}                    &  \colhead{(min)}         &  \colhead{(mm)}                             & \colhead{Antennas} & \colhead{(m)} & \colhead{(m)}   &  \colhead{($\arcsec$ $\times$ $\arcsec$)} &  \colhead{($\arcsec$)}                       &  \colhead{}                }
\startdata 
Band 6       &  2016 Nov 30               &  16.1                          &  1.9--2.5                                         &  44                            &  15.1               &  704.1              &  0.54 $\times$ 0.40                                     &   3.6                              &  0.98 MHz               \\
(250 GHz)  &  (Cycle 4)                     &                                   &                                                       &                                  &  \multicolumn{2}{c}{(C40-4)}      &                                                                     &                                      &  (1.2 km s$^{-1}$)    \\
Band 7       &  2018 Dec 4                 &  35.8                          &  0.5--0.6                                         &  46                            &  15.1               &  783.5              &  0.37 $\times$ 0.32                                     &   3.3                              &  0.98 MHz                \\
(350 GHz)  &  (Cycle 6)                     &                                   &                                                       &                                  &  \multicolumn{2}{c}{(C43-4)}      &                                                                     &                                      &  (0.85 km s$^{-1}$)    \\
\enddata
\tablecomments{
$^a$Precipitable water vapor. 
$^b$The average beam size achieved by TCLEAN with the Briggs weighting and the robustness parameter of 0.5. 
Note that we use a common circular restoring beam size of 0$\farcs$40 for Band 6 and 7 data to construct the final images. 
$^c$Maximum Recoverable Scale. 
} 
\end{deluxetable*}

\section{Target, observations, and data reduction} \label{sec_tarobsred} 
\subsection{Target} \label{sec_tar}
\begin{figure}[tp]
\begin{center}
\includegraphics[width=8.5cm]{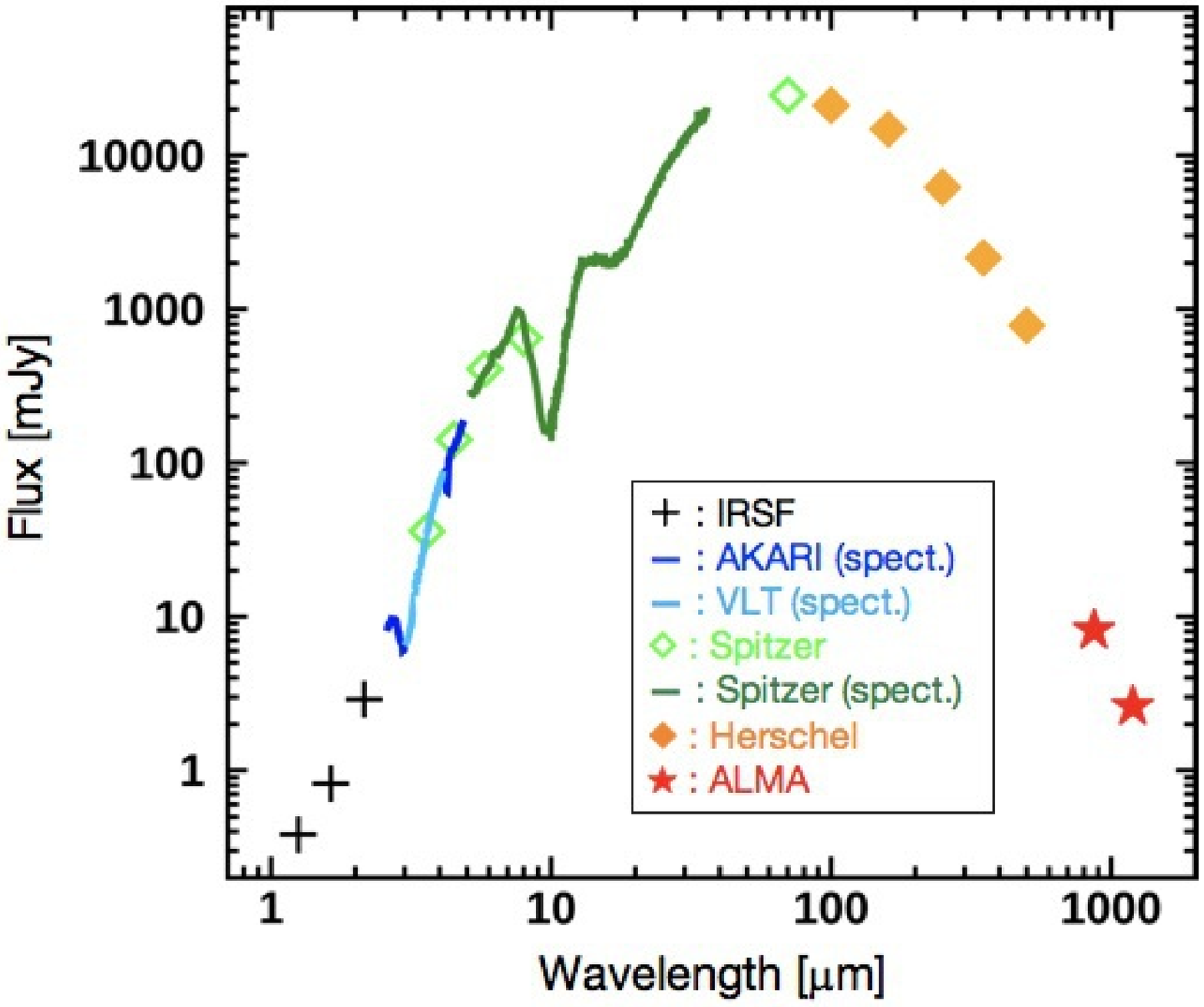}
\caption{
The spectral energy distribution of the observed high-mass young stellar object, ST16. 
The plotted data are based on 
IRSF/SIRIUS JHK$_\mathrm{s}$ photometry \citep[pluses, black,][]{Kat07}, 
{\it AKARI}/IRC spectroscopy \citep[solid line, blue,][]{ST10}, 
VLT/ISAAC spectroscopy \citep[solid line, light blue,][]{ST16}, 
{\it Spitzer}/IRAC and MIPS photometry \citep[open diamonds, light green,][]{Mei06}, 
{\it Spitzer}/MIPS spectroscopy \citep[solid line, green,][]{Kem10}, 
{\it Herschel}/PACS and SPIRE photometry \citep[filled diamonds, orange,][]{Mei13}, 
and ALMA 870 $\mu$m and 1200 $\mu$m continuum measurements obtained in this work (filled star, red). 
}
\label{sed}
\end{center}
\end{figure}

The target of the present ALMA observations is an infrared source, IRAS 05195-6911 or ST16 (hereafter ST16), located near the N119 star-forming region in the LMC.  
Previous infrared spectroscopic studies have classified the source as an embedded high-mass YSO \citep{Sea09,ST16}. 
A spectral energy distribution (SED) of the source is shown in Figure \ref{sed} \citep[data are collected from available databases and literatures including][]{Mei06, Mei13, Kat07, ST16, Kem10}. 
The bolometric luminosity of the source is estimated to be 3.1 $\times$ 10$^5$ L$_{\sun}$ by integrating the SED from 1 $\mu$m to 1200 $\mu$m.

\subsection{Observations} \label{sec_obs} 
Observations were carried out with ALMA as a part of Cycle 4 (2016.1.00394.S) and Cycle 6 (2018.1.01366.S) programs (PI T. Shimonishi). 
A summary of the present observations is shown in Table \ref{tab_Obs}. 
The target high-mass YSO is located at RA = 05$^\mathrm{h}$19$^\mathrm{m}$12$\fs$31 and Dec = -69$^\circ$9$\arcmin$7$\farcs$3 (ICRS), based on the {\it Spitzer} SAGE infrared catalog \citep{Mei06}. 
The source's positional accuracy is about 0.3$\arcsec$. 
The pointing center of antennas is RA = 05$^\mathrm{h}$19$^\mathrm{m}$12$\fs$30 and Dec = -69$^\circ$9$\arcmin$6$\farcs$8 (ICRS), which roughly corresponds to the infrared center of the target. 

The total on-source integration time is 16.1 minutes for Band 6 data and 35.8 minutes for Band 7. 
Flux, bandpass, and phase calibrators are J0519-4546, J0635-7516, and J0526-6749 for Band 6, while J0519-4546, J0519-4546, and J0529-7245 for Band 7, respectively. 
Four spectral windows are used to cover the sky frequencies of 241.25--243.12, 243.61-245.48, 256.75--258.62, and 258.60--260.48 GHz for Band 6 and 336.97--338.84, 338.77-340.64, 348.85--350.72, and 350.65--352.53 GHz for Band 7. 
The channel spacing is 0.98 MHz, which corresponds to 1.2 km s$^{-1}$ for Band 6 and 0.85 km s$^{-1}$ for Band 7. 
The total number of antennas is 44 for Band 6 and 46 for Band 7. 
The minimum--maximum baseline lengths are 15.1--704.1 m for Band 6 and 15.1--783.5 m for Band 7. 
A full-width at half-maximum (FWHM) of the primary beam is about 25$\arcsec$ for Band 6 and 18$\arcsec$ for Band 7.

\subsection{Data reduction} \label{sec_red} 
Raw data are processed with the \textit{Common Astronomy Software Applications} (CASA) package. 
For calibration, CASA 4.7.2 is used for Band 6 and CASA 5.4.0 is used for Band 7. 
For imaging, we use CASA 5.4.0 for the all data. 
With the Briggs weighting and the robustness parameter of 0.5, the synthesized beam sizes of 0$\farcs$52--0$\farcs$56 $\times$ 0$\farcs$39--0$\farcs$41 with a position angle of -23 degree for Band 6 and 0$\farcs$36--0$\farcs$38 $\times$ 0$\farcs$31--0$\farcs$32 with a position angle of -17 for Band 7 are achieved. 
In this paper, we have used a common circular restoring beam size of 0$\farcs$40 for Band 6 and 7 data, in order to accommodate the spectral analyses in separated frequency regions. 
This beam size corresponds to 0.097 pc at the distance of the LMC. 
The continuum image is constructed by selecting line-free channels from the four spectral windows. 
After the clean process, the images are corrected for the primary beam pattern using the \textit{impbcor} task in CASA. 
The self-calibration is not applied. 

The spectra and continuum flux are extracted from the 0$\farcs$42 (0.10 pc) diameter circular region centered at RA = 05$^\mathrm{h}$19$^\mathrm{m}$12$\fs$295 and Dec = -69$^\circ$9$\arcmin$7$\farcs$34 (ICRS), which corresponds to the 870 $\mu$m continuum center of ST16. 
The continuum emission is subtracted from the spectral data using the \textit{uvcontsub} task in CASA before the spectral extraction.

\section{Results} \label{sec_res} 
\subsection{Spectra} \label{sec_spc} 
\begin{figure*}[ht]
\centering
\includegraphics[width=21cm,angle=90]{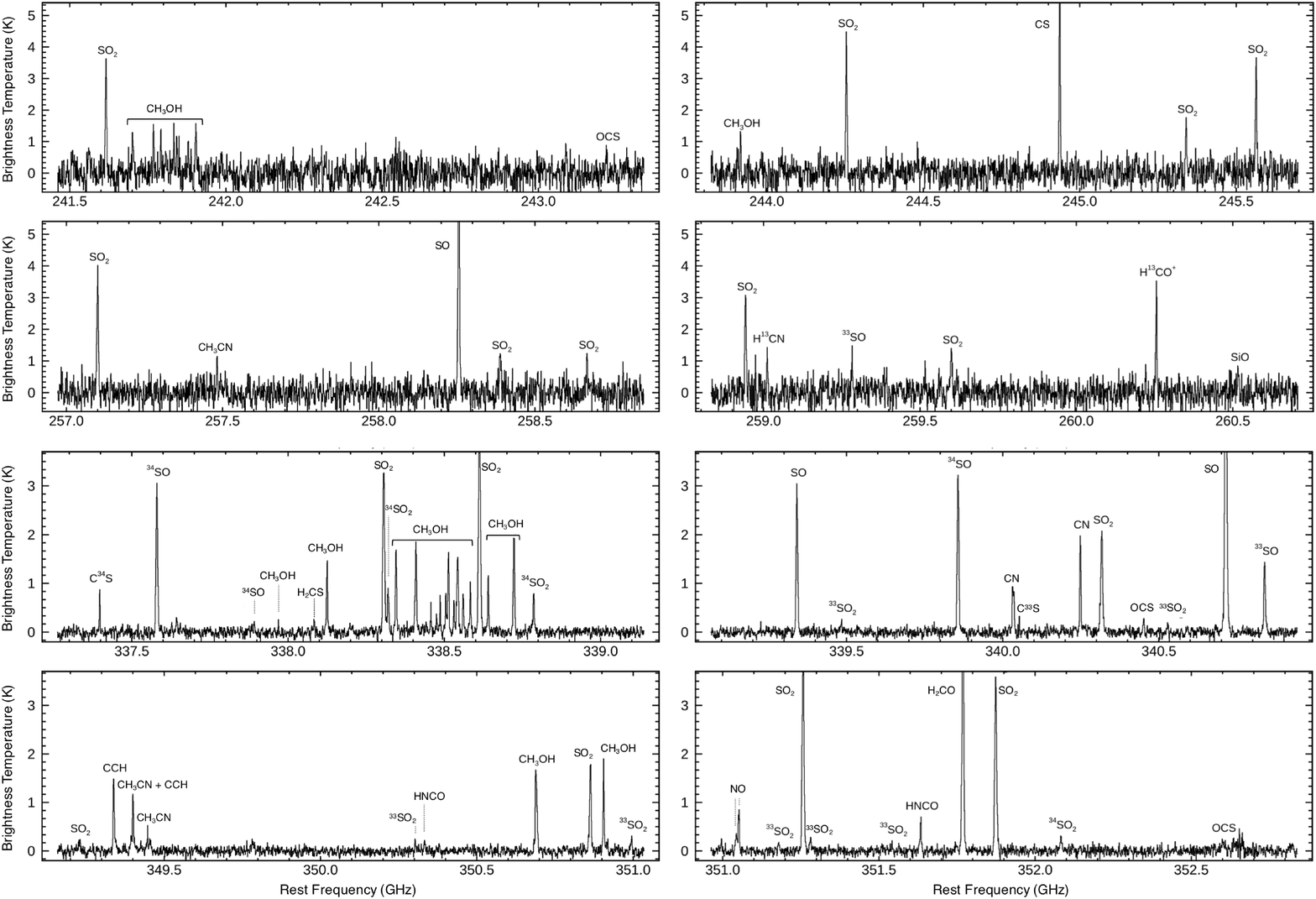}
\caption{ALMA Band 6 and 7 spectra of ST16 extracted from a 0$\farcs$42 (0.10 pc) diameter region centered at the continuum and molecular emission peak. 
Detected emission lines are labeled. 
The source velocity of 264.5 km s$^{-1}$ is assumed. 
}
\label{spec}
\end{figure*}

Figure \ref{spec} shows molecular emission line spectra extracted from the position of ST16. 
Spectral lines are identified with the aid of the Cologne Database for Molecular Spectroscopy\footnote{https://www.astro.uni-koeln.de/cdms} \citep[CDMS,][]{Mul01,Mul05} and the molecular database of the Jet Propulsion Laboratory\footnote{http://spec.jpl.nasa.gov} \citep[JPL,][]{Pic98}. 
The detection criteria adopted here are the 2.5$\sigma$ significance level and the velocity coincidence with the systemic velocity of nearby CO clouds \citep[between 260 km s$^{-1}$ and 270 km s$^{-1}$, estimated using the MAGMA data presented in][]{Won11}. 

Molecular emission lines of CH$_3$OH, H$_2$CO, CCH, H$^{13}$CO$^{+}$, CS, C$^{34}$S, C$^{33}$S, SO, $^{34}$SO, $^{33}$SO, SO$_2$, $^{34}$SO$_2$, $^{33}$SO$_2$, OCS, H$_2$CS, CN, NO, HNCO, H$^{13}$CN, CH$_3$CN, and SiO are detected from the observed region. 
Multiple high excitation lines (upper state energy $>$100 K) are detected for CH$_3$OH, SO$_2$, $^{34}$SO$_2$, $^{33}$SO$_2$, OCS, and CH$_3$CN. 
Complex organic molecules larger than CH$_3$OH are not detected. 
In total we have detected 90 transitions, out of which, 30 lines are due to CH$_3$OH, and 27 lines are due to SO$_2$ and its isotopologues. 
Radio recombination lines are not detected, though moderately-intense lines such as H36$\beta$ (260.03278 GHz) or H41$\gamma$ (257.63549 GHz) are covered in the observed frequency range. 

Line parameters are measured by fitting a Gaussian profile to observed lines. 
Based on the fitting, we estimate the peak brightness temperature, the FWHM, the LSR velocity, and the integrated intensity for each line. 
Measured line widths are typically 3--6 km s$^{-1}$. 
Full details of the line fitting can be found in Appendix A (Tables of measured line parameters) and Appendix B (Figures of fitted spectra).
The tables also contain estimated upper limits on important non-detection lines.

\subsection{Images} \label{sec_img} 
\begin{figure*}[tp]
\begin{center}
\includegraphics[width=15.0cm]{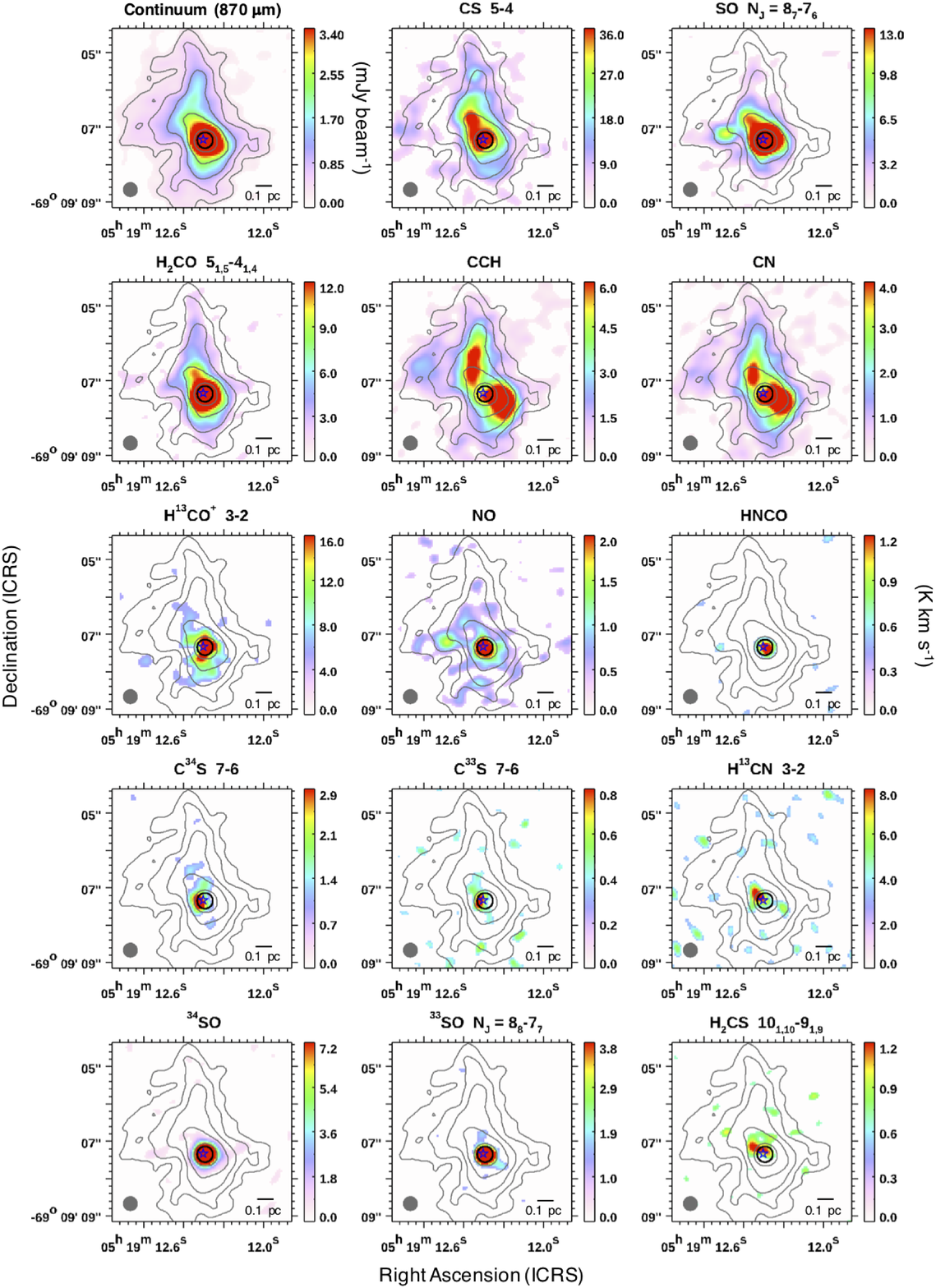}
\caption{
Flux distributions of the ALMA 870 $\mu$m continuum and integrated intensity distributions of molecular emission lines. 
For CCH, CN, NO, HNCO, and $^{34}$SO, the detected multiple transitions are averaged. 
Gray contours represent the continuum distribution and the contour levels are 5$\sigma$, 10$\sigma$, 20$\sigma$, 40$\sigma$, 100$\sigma$ of the rms noise (0.06 mJy/beam). 
Low signal-to-noise regions (S/N $<$2) are masked. 
The spectra discussed in the text are extracted from the region indicated by the thick black open circle. 
The blue open star represents the position of a high-mass YSO identified by infrared observations. 
The synthesized beam size is shown by the gray filled circle in each panel. 
North is up, and east is to the left. 
}
\label{images1}
\end{center}
\end{figure*}

\begin{figure*}[p]
\begin{center}
\includegraphics[width=15.0cm]{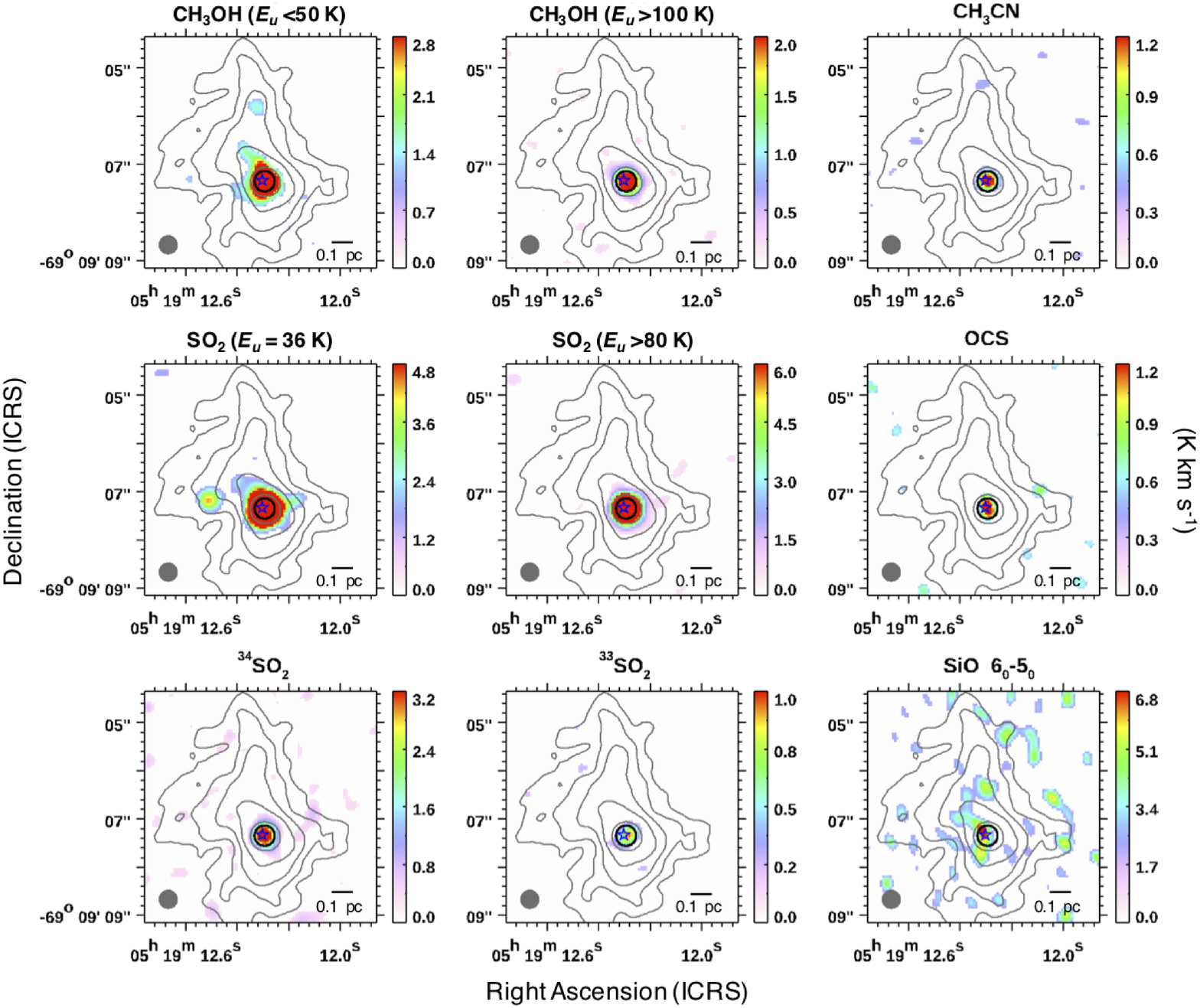}
\caption{
Same as in Figure \ref{images1}. 
For CH$_3$OH, CH$_3$CN, SO$_2$, OCS, $^{34}$SO$_2$, $^{33}$SO$_2$, the detected multiple transitions are averaged. 
CH$_3$OH and SO$_2$ are separated into high-$E_{u}$ ($>$100 K for CH$_3$OH and $>$80 K for SO$_2$) and low-$E_{u}$ ($<$50 K for CH$_3$OH and $=$ 36 K for SO$_2$) components. 
}
\label{images2}
\end{center}
\end{figure*}

Figures \ref{images1} and \ref{images2} shows synthesized images of continuum and molecular emission lines observed toward the target region. 
The molecular line images are constructed by integrating spectral data in the velocity range where the emission is seen. 
For CH$_3$OH and SO$_2$, high-excitation line images ($E_{u}$ $>$100 K for CH$_3$OH and $>$80 K for SO$_2$) and low-excitation line images ($E_{u}$ $<$50 K for CH$_3$OH and 36 K for SO$_2$) are separately constructed, because these molecules show two different temperature components in their rotation diagrams (see Section \ref{sec_rd}). 

The continuum emission, as well as most of molecular emission lines except for CCH and CN, are centered at the position of the high-mass YSO. 
SO, NO, and low-$E_{u}$ SO$_2$ show a secondary peak at the east side of the YSO. 
The distributions of continuum, CS, SO, H$_2$CO, CCH, CN, and possibly H$^{13}$CO$^{+}$ are elongated in the north-south direction. 

We have estimated the spatial extent of each emission around the YSO by fitting a two-dimensional Gaussian to the peak position. 
Compact distributions, i.e. FWHM = 0$\farcs$38--0$\farcs$46 (0.09--0.11 pc) that is comparable with the beam size, are seen in high-$E_{u}$ CH$_3$OH, low- and high-$E_{u}$ SO$_2$, $^{34}$SO$_2$, $^{33}$SO$_2$, OCS, $^{34}$SO, $^{33}$SO, CH$_3$CN, and HNCO. 
Slightly extended distributions, i.e. FWHM = 0$\farcs$52--0$\farcs$76 (0.13--0.18 pc), are seen in H$_2$CO, low-$E_{u}$ CH$_3$OH, SO, C$^{34}$S, C$^{33}$S, H$_2$CS, NO, H$^{13}$CN, SiO, and continuum. 
Among them, the distributions of H$^{13}$CN and H$_2$CS, and possibly C$^{34}$S, C$^{33}$S, and SiO, are marginally off the continuum center. 
The NO distribution seems to be patchy. 
The continuum emission has a sharp peak around the YSO position, but also widely distribute within the observed field as shown by the 5$\sigma$ contour in Figure \ref{images1}. 
Similar characteristics (sharp peak and extended plateau) are also seen SO and H$_2$CO. 
Clearly extended distributions, i.e. FWHM = 1$\arcsec$--2$\arcsec$ (0.24--0.49 pc), are seen in CS, CCH, CN, and H$^{13}$CO$^{+}$. 
The distributions of CCH and CN are significantly different from those of other molecules and will be further discussed in Section \ref{sec_disc_CCH_CN}. 

\section{Analysis} \label{sec_ana} 
\subsection{Rotation diagram analyses} \label{sec_rd}
\begin{figure*}[tp]
\begin{center}
\includegraphics[width=18cm]{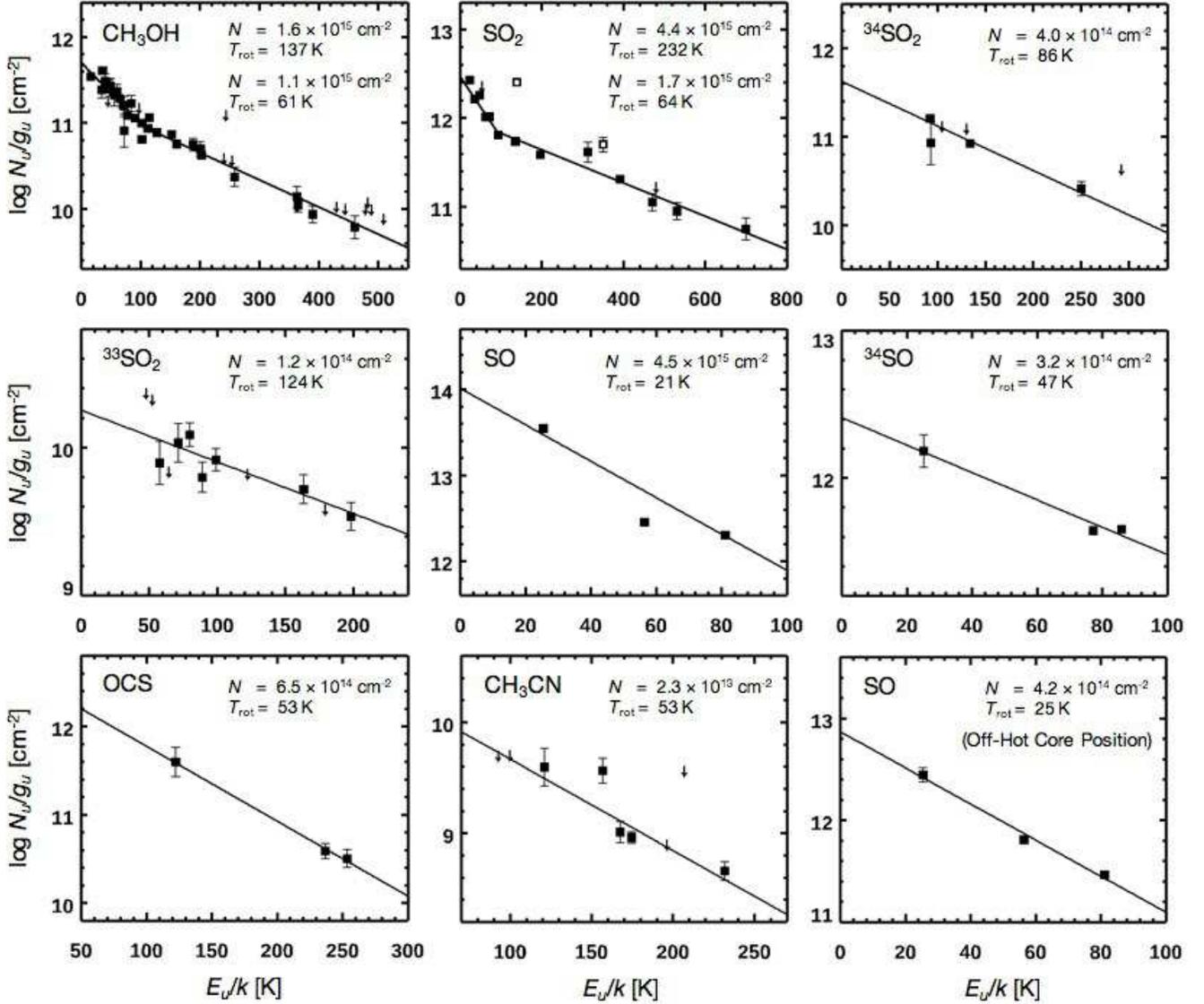}
\caption{
Rotation diagrams for CH$_3$OH, SO$_2$, $^{34}$SO$_2$, $^{33}$SO$_2$, SO, $^{34}$SO, OCS, and CH$_3$CN lines. 
Upper limit points are shown by the downward arrows. 
The solid lines represent the fitted straight line. 
Derived column densities and rotation temperatures are shown in each panel. 
For CH$_3$OH and SO$_2$, the left line is fitted using the transitions with $E_{u}$ $<$100 K, while the right one is fitted using the transitions with $E_{u}$ $>$100 K. 
Note that A- and E-state CH$_3$OH are fitted simultaneously. 
Two SO$_2$ transitions, 10$_{6,4}$--11$_{5,7}$ and 26$_{3,23}$--25$_{4,22}$ (indicated by the open squares), are excluded from the fit, because they significantly deviates from other data points. 
At the lower-right panel, the rotation diagram of SO at the off-hot core position is shown. 
See Section \ref{sec_rd} for details. 
}
\label{rd}
\end{center}
\end{figure*}

Column densities and rotation temperatures of CH$_3$OH, SO$_2$, $^{34}$SO$_2$, $^{33}$SO$_2$, SO, $^{34}$SO, OCS, and CH$_3$CN are estimated with the aid of the rotation diagram analysis, because multiple lines with different excitation energies are detected (Figure \ref{rd}). 
We here assume an optically thin condition and the local thermodynamic equilibrium (LTE). 
The assumption of optically thin emission is mostly valid for the present source (see discussion in Sections \ref{sec_disc_molab}, \ref{sec_disc_isotop}, and \ref{sec_disc_CCH_CN}). 
We use the following formulae based on the standard treatment of the rotation diagram analysis \citep[e.g., ][]{Sut95, Gol99}: 
\begin{equation}
\log \left(\frac{ N_{u} }{ g_{u} } \right) = - \left(\frac {\log e}{T_{\mathrm{rot}}} \right) \left(\frac{E_{u}}{k} \right) + \log \left(\frac{N}{Q(T_{\mathrm{rot}})} \right),  \label{Eq_rd1}
\end{equation}
where 
\begin{equation}
\frac{ N_{u} }{ g_{u} } = \frac{ 3 k \int T_{\mathrm{b}} dV }{ 8 \pi^{3} \nu S \mu^{2} }, \label{Eq_rd2} \\ 
\end{equation}
and $N_{u}$ is a column density of molecules in the upper energy level, $g_{u}$ is the degeneracy of the upper level, $k$ is the Boltzmann constant, $\int T_{\mathrm{b}} dV$ is the integrated intensity estimated from the observations, $\nu$ is the transition frequency, $S$ is the line strength, $\mu$ is the dipole moment, $T_{\mathrm{rot}}$ is the rotational temperature, $E_{u}$ is the upper state energy, $N$ is the total column density, and $Q(T_{\mathrm{rot}})$ is the partition function at $T_{\mathrm{rot}}$. 
All the spectroscopic parameters required in the analysis are extracted from the CDMS database. 

For CH$_3$OH and SO$_2$, a straight-line fit is separated into two temperature regimes, because different temperature components are clearly seen in the diagram. 
We use the transitions with $E_{u}$ $<$100 K to fit the lower-temperature component and the transitions with $E_{u}$ $>$100 K to fit the higher-temperature component, respectively. 
Derived column densities and rotation temperatures are summarized in Table \ref{tab_N}. 
These rotation analyses suggest that the line-of-sight towards ST16 harbors two temperature components; i.e. a hot gas component with $T_{\mathrm{rot}}$ $\sim$150 K (an average temperature of high-temperature CH$_3$OH, high-temperature SO$_2$, $^{34}$SO$_2$, and $^{33}$SO$_2$), along with a warm gas component with $T_{\mathrm{rot}}$ $\sim$50 K (an average temperature of low-temperature CH$_3$OH, low-temperature SO$_2$, $^{34}$SO, OCS, and CH$_3$CN). 

Note that the temperature and column density derived from the rotation diagram of SO would not be reliable, because the SO(6$_{6}$--5$_{5}$) and (8$_{7}$--7$_{6}$) lines are moderately optically thick ($\tau$ $\sim$0.3--1, with $T_{\mathrm{rot}}$ = 50--20 K and a beam filling factor of unity). 
Given a possible beam dilution effect for high excitation lines, their optical thickness would cause non-negligible uncertainty on the reliability of the rotation analysis. 
We thus derive the SO column density using the SO(3$_{3}$--2$_{3}$) line, which has an $S \mu^{2}$ value about hundred times smaller than the above two transitions and is optically thin. 
Here the rotation temperature is assumed to be the same as that of $^{34}$SO ($\sim$50 K). 

We have also estimated the rotational temperature of SO at the off-hot-core position (i.e. the 0$\farcs$40 diameter circular region centered at RA = 05$^\mathrm{h}$19$^\mathrm{m}$12$\fs$39 and Dec = -69$^\circ$9$\arcmin$6$\farcs$6). 
Here we derive $T_{\mathrm{rot}}$ = 24.5 $\pm$ 1.4 K and $N$ = 4.2 $\pm$ 0.6 $\times$ 10$^{14}$ cm$^{-2}$ (Figure \ref{rd}, lower right panel). 
Note that the SO lines at the off position are optically thin ($<$0.15), since the peak intensities are nearly five times lower than those at the hot core position. 
The derived $T_{\mathrm{rot}}$ would represent the temperature of the relatively cold and dense gas surrounding the hot core.

\subsection{Column densities of other molecules} \label{sec_n} 
Column densities of molecular species other than those described in Section \ref{sec_rd} are estimated from Equation \ref{Eq_rd1} after solving it for $N$. 
For this purpose we need to assume their rotation temperatures. 

The rotation temperature of $^{34}$SO ($\sim$50K) corresponds to the temperature of warm components in the line of sight. 
This temperature is applied to SO and $^{33}$SO considering the co-existence of isotopologues. 
CS shows an extended distribution similar to SO, we thus also applied $T_{\mathrm{rot}}$ = 50 K for CS, C$^{34}$S, and C$^{33}$S. 
Furthermore, we also assume $T_{\mathrm{rot}}$ = 50 K for CCH and CN, because they are clearly extended and not centered at the hot core region. 
For other molecules that are concentrated at the hot core, we assume $T_{\mathrm{rot}}$ = 100 K, which is an average temperature of the hot and warm gas components describe in Section \ref{sec_rd}. 

We use the spectroscopic constants and partition functions extracted from the CDMS database except for HCOOCH$_3$, whose molecular data is extracted from the JPL database. 
Estimated column densities are summarized in Table \ref{tab_N}. 

We have also performed non-LTE calculation for column densities of selected species using RADEX \citep{vdT07}. 
For input parameters, we use the H$_2$ gas density of 3 $\times$ 10$^6$ cm$^{-3}$ according to our estimate in Section \ref{sec_h2_final} and the background temperature of 2.73 K. 
Kinetic temperatures are assumed to be the same as temperatures tabulated in Table \ref{tab_N}. 
The line widths used in the analysis are taken from the tables in Appendix A. 
The resultant column densities are summarized in Table \ref{tab_N}. 
The calculated non-LTE column densities are reasonably consistent with the LTE estimations.

\begin{deluxetable}{ l c c c}
\tablecaption{Estimated rotation temperatures and column densities \label{tab_N}}
\tablewidth{0pt}
\tabletypesize{\footnotesize} 
\tablehead{
\colhead{Molecule}   & \colhead{$T$$_{rot}$}   &       \colhead{$N$(X)}            &  \colhead{$N$(X) non-LTE\tablenotemark{d}}  \\
\colhead{ }                & \colhead{(K)}                 &        \colhead{(cm$^{-2}$)}         & \colhead{(cm$^{-2}$)} 
}
\startdata 
 CH$_3$OH\tablenotemark{a} ($E_{u} >$100 K) &   137$^{+8}_{-7}$    &    (1.6 $^{+0.1}_{-0.1}$) $\times$ 10$^{15}$     & (3.4 $\pm$ 0.2) $\times$ 10$^{15}$ \\
 CH$_3$OH\tablenotemark{a} ($E_{u} <$100 K) &   61$^{+2}_{-2}$      &    (1.1 $^{+0.1}_{-0.1}$) $\times$ 10$^{15}$     & (1.6 $\pm$ 0.1) $\times$ 10$^{15}$  \\
 SO$_2$\tablenotemark{a} ($E_{u} >$100 K)       &   232$^{+27}_{-22}$   &     (4.4 $^{+0.1}_{-0.1}$) $\times$ 10$^{15}$ & (3.5 $\pm$ 0.3) $\times$ 10$^{15}$    \\
 SO$_2$\tablenotemark{a} ($E_{u} <$100 K)      &     64$^{+5}_{-4}$    &     (1.7 $^{+0.1}_{-0.1}$) $\times$ 10$^{15}$     & (2.0 $\pm$ 0.1) $\times$ 10$^{15}$ \\
 $^{34}$SO$_2$\tablenotemark{a}                    &     86$^{+11}_{-9}$    &     (4.0 $^{+0.8}_{-0.6}$) $\times$ 10$^{14}$      & \nodata \\
 $^{33}$SO$_2$\tablenotemark{a}                    &     124$^{+37}_{-23}$    &     (1.2 $^{+0.3}_{-0.2}$) $\times$ 10$^{14}$  & \nodata   \\
$^{34}$SO\tablenotemark{a,b}                            &     47$^{+14}_{-14}$    &     (3.2 $^{+0.9}_{-0.9}$) $\times$ 10$^{14}$  & \nodata   \\
OCS\tablenotemark{a}                                  &     53$^{+11}_{-8}$    &     (6.5 $^{+7.2}_{-3.4}$) $\times$ 10$^{14}$          & (6.9 $\pm$ 0.5) $\times$ 10$^{14}$  \\
CH$_3$CN\tablenotemark{a}                           &     53$^{+10}_{-7}$    &     (2.3 $^{+1.7}_{-1.0}$) $\times$ 10$^{13}$      & (2.5 $\pm$ 0.2) $\times$ 10$^{13}$ \\
\tableline
SO\tablenotemark{c}             &   50        &  (7.3 $\pm$ 0.2) $\times$ 10$^{15}$     & (7.5 $\pm$ 0.1) $\times$ 10$^{15}$ \\
$^{33}$SO                               &  50        &   (1.4 $\pm$ 0.1) $\times$ 10$^{14}$   & \nodata  \\
CS                                          &   50        &  (1.3 $\pm$ 0.1) $\times$ 10$^{14}$    & (1.2 $\pm$ 0.1) $\times$ 10$^{14}$  \\
C$^{34}$S                               &   50        &   (6.7 $\pm$ 0.4) $\times$ 10$^{12}$  &  \nodata  \\
C$^{33}$S                               &   50        &   (1.8 $\pm$ 0.3) $\times$ 10$^{12}$   & \nodata \\
CCH                                       &   50        &   (1.3 $\pm$ 0.1) $\times$ 10$^{14}$   &  \nodata  \\
CN                                          &   50        &   (4.5 $\pm$ 0.5) $\times$ 10$^{13}$   &  \nodata  \\
H$_2$CO                                &   100        &   (2.1 $\pm$ 0.1) $\times$ 10$^{14}$  & (1.3 $\pm$ 0.1) $\times$ 10$^{14}$  \\
H$^{13}$CO$^+$                     &   100        &   (1.4 $\pm$ 0.1) $\times$ 10$^{13}$ & (7.7 $\pm$ 0.3) $\times$ 10$^{12}$  \\
NO                                         &   100        &   (4.1 $\pm$ 0.8) $\times$ 10$^{15}$   & (3.8 $\pm$ 0.2) $\times$ 10$^{15}$ \\ 
HNCO                                     &   100        &   (3.7 $\pm$ 0.8) $\times$ 10$^{13}$  & (2.3 $\pm$ 0.2) $\times$ 10$^{13}$  \\ 
H$^{13}$CN                             &   100        &   (8.8 $\pm$ 2.0) $\times$ 10$^{12}$  &  (4.2 $\pm$ 0.2) $\times$ 10$^{12}$ \\
HC$_3$N                                &   100        &   $<$1.1 $\times$ 10$^{13}$               & \nodata  \\
H$_2$CS                                &   100        &   (1.9 $\pm$ 0.4) $\times$ 10$^{13}$  & (1.3 $\pm$ 0.2) $\times$ 10$^{13}$   \\
SiO                                        &   100        &   (8.5 $\pm$ 3.2) $\times$ 10$^{12}$   & (6.0 $\pm$ 1.0) $\times$ 10$^{12}$   \\
HDO                                      &   100        &    $<$3.8 $\times$ 10$^{14}$   & \nodata   \\
c-C$_3$H$_2$                        &    100       &   $<$2.4 $\times$ 10$^{14}$  &  \nodata  \\
C$_2$H$_5$OH                      &   100        &   $<$4.1 $\times$ 10$^{14}$  &  \nodata  \\ 
C$_2$H$_5$CN                      &   100        &   $<$4.5 $\times$ 10$^{14}$  &  \nodata  \\
CH$_3$OCH$_3$                   &   100        &   $<$2.5 $\times$ 10$^{14}$  & \nodata   \\  
HCOOCH$_3$                        &   100        &   $<$6.8 $\times$ 10$^{14}$  & \nodata   \\
\textit{trans}-HCOOH                &   100        &  $<$7.5 $\times$ 10$^{13}$  &  \nodata \\ 
\enddata
\tablecomments{
Uncertainties and upper limits are of the 2 $\sigma$ level and do not include systematic errors due to adopted spectroscopic constants. 
See Section \ref{sec_rd} and \ref{sec_n} for details.  
$^a$Derived based on the rotation diagram analysis. 
$^b$Assumed empirical 30 $\%$ uncertainty for $T$$_{rot}$ and $N$ because of the fitted data points are relatively few and scattered. 
$^c$Derived from the SO(3$_{3}$--2$_{3}$) line. 
$^d$The following lines are used for non-LTE calculation with RADEX; 
CH$_3$OH(7$_{2}$ A$^+$--6$_{2}$ A$^+$), CH$_3$OH(5$_{1}$ E--4$_{1}$ E), SO$_2$(5$_{3,3}$--4$_{2,2}$), SO$_2$(14$_{0,14}$--13$_{1,13}$), OCS(28--27), CH$_3$CN(19$_{0}$--18$_{0}$), SO($N_J$ = 6$_{6}$--5$_{5}$), CS(5--4), H$_2$CO(5$_{1,5}$--4$_{1,4}$), H$^{13}$CO$^+$(3--2), NO(J = $\frac{7}{2}$--$\frac{5}{2}$, $\Omega$ = $\frac{1}{2}$, F = $\frac{9}{2}$$^+$--$\frac{7}{2}$$^-$), HNCO(16$_{0,16}$--15$_{0,15}$), H$^{13}$CN(3--2), H$_2$CS(10$_{1,10}$--9$_{1,9}$), and SiO(6--5). 
}
\end{deluxetable}

\subsection{Column density of H$_2$} \label{sec_h2} 
A column density of molecular hydrogen ($N_{\mathrm{H_2}}$) is estimated by several methods to check their reliability. 
The H$_2$ column densities derived by different methods are summarized in Table \ref{tab_N_H2}. 
Details of each method are described below.

\begin{deluxetable*}{ l c c c c c c c c c c} 
\tablecaption{Estimated H$_2$ column densities and $A_{V}$ \label{tab_N_H2}} 
\tablewidth{0pt} 
\tabletypesize{\scriptsize} 
\tablehead{
\colhead{ }                & \multicolumn{6}{c}{ALMA continuum} & \colhead{ } & \multicolumn{2}{c}{SED fit}  &  \colhead{$\tau$$_{9.7}$}  \\
\cline{2-7}
\cline{9-10}
\colhead{ }                & \multicolumn{2}{c}{$T_{d}$ = 20 K}  & \multicolumn{2}{c}{$T_{d}$ = 60 K}  & \multicolumn{2}{c}{$T_{d}$ = 150 K}  & \colhead{ } & \colhead{RW07\tablenotemark{a}} & \colhead{ZT18\tablenotemark{b}} &  \colhead{ } \\
\cline{2-3}
\cline{4-5}
\cline{6-7}
\colhead{ }                & \colhead{870 $\mu$m}  & \colhead{1200 $\mu$m}  & \colhead{870 $\mu$m}  & \colhead{1200 $\mu$m}  & \colhead{870 $\mu$m}  & \colhead{1200 $\mu$m}  & \colhead{ }  & \colhead{}   & \colhead{ }           & \colhead{} 
}
\startdata 
$N_{\mathrm{H_2}}$ (10$^{23}$ cm$^{-2}$)  &  21.0 $\pm$ 2.1     &  22.2 $\pm$ 2.2      &    5.64 $\pm$ 0.56     &  5.47 $\pm$ 0.55     &  2.12 $\pm$ 0.21        &  2.01 $\pm$ 0.20         &  &  5.40 $\pm$ 1.12 &  5.71 $\pm$ 0.87  &  5.60 $\pm$ 0.56  \\
$A_{V}$ (mag)                                          &   749 $\pm$ 75       &  792 $\pm$ 79         &    202 $\pm$ 20        & 195 $\pm$ 20         &   76 $\pm$ 8              &  72 $\pm$ 7                 &  &  193 $\pm$ 40     &  204 $\pm$ 31     &  200 $\pm$ 20       \\
\enddata
\tablecomments{
In this work, we use $N_{\mathrm{H_2}}$ = (5.6 $\pm$ 0.6) $\times$ 10$^{23}$ cm$^{-2}$ as a representative value, which is the average of $N_{\mathrm{H_2}}$ derived by ALMA dust continuum (870 $\mu$m and 1200 $\mu$m with $T_{d}$ = 60 K), the SED model fits, and the 9.7 $\mu$m silicate dust absorption depth (see Section \ref{sec_h2} for details). 
Uncertainties do not include systematic errors due to adopted optical constants. 
\\
$^a$\citet{Rob07}; 
$^b$\citet{Zha18}
} 
\end{deluxetable*}

\subsubsection{$N_{\mathrm{H_2}}$ from the ALMA continuum} \label{sec_h2_dust} 
The present ALMA dust continuum data can be used for the $N_{\mathrm{H_2}}$ estimate. 
The continuum brightness of ST16 is measured to be (3.37 $\pm$ 0.34) mJy/beam for 1200 $\mu$m and (10.55 $\pm$ 1.06) mJy/beam for 870 $\mu$m towards the region same as in the spectral extraction\footnote{A canonical uncertainty of 10 $\%$ for the absolute flux calibration of the ALMA Band 6 and 7 data is adopted (see ALMA Technical Handbook).}
Based on the standard treatment of optically thin dust emission, we use the following equation to calculate $N_{\mathrm{H_2}}$: 
\begin{equation}
N_{\mathrm{H_2}} = \frac{F_{\nu} / \Omega}{2 \kappa_{\nu} B_{\nu}(T_{d}) Z \mu m_{\mathrm{H}}} \label{Eq_h2}, 
\end{equation}
where $F_{\nu}/\Omega$ is the continuum flux density per beam solid angle as estimated from the observations, 
$\kappa_{\nu}$ is the mass absorption coefficient of dust grains coated by thin ice mantles at 1200/870 $\mu$m as taken from \citet{Oss94} and we here use 1.06 cm$^2$ g$^{-1}$ for 1200 $\mu$m and 1.89 cm$^2$ g$^{-1}$ for 870 $\mu$m, 
$T_{d}$ is the dust temperature and $B_{\nu}(T_{d})$ is the Planck function, 
$Z$ is the dust-to-gas mass ratio, $\mu$ is the mean atomic mass per hydrogen \citep[1.41, according to][]{Cox00}, and $m_{\mathrm{H}}$ is the hydrogen mass. 
We use the dust-to-gas mass ratio of 0.0027 for the LMC, which is obtained by scaling the Galactic value of 0.008 by the metallicity of the LMC ($\sim$1/3 Z$_{\sun}$). 

The dust temperature is a key assumption for the derivation of $N_{\mathrm{H_2}}$. 
We estimate $N_{\mathrm{H_2}}$ for three different dust temperatures, 20 K, 60 K, and 150 K, as shown in Table \ref{tab_N_H2}. 
We revisit the validity of these assumption in Section \ref{sec_h2_final}, based on the comparison of $N_{\mathrm{H_2}}$ values by different methods. 
Note that consistent $N_{\mathrm{H_2}}$ values are derived from 870 $\mu$m and 1200 $\mu$m continuum, suggesting that the submillimeter continuum emission from ST16 is almost dominated by the thermal emission from dust grains.

\subsubsection{$N_{\mathrm{H_2}}$ from the SED fit} \label{sec_h2_sed} 
A model fit to the source's SED provides us with an alternative way to estimate the total gas column density in the line-of-sight. 
We have tested two SED models in this work; one by \citet{Rob07} and another by \citet{Zha18}. 
For input data, we use 1--1200 $\mu$m photometric and spectroscopic data of ST16 as shown in Figure \ref{sed}. 
We exclude the SPIRE 350 $\mu$m and 500 $\mu$m band data in the fit, because they are possibly contaminated by diffuse emission around the YSO due to their large point spread function (about 27$\arcsec$ and 41$\arcsec$ in FWHM, respectively). 
The distance to ST16 is assumed to be the same as that of the LMC. 

The model of \citet{Rob07} produces a bunch of fitted SEDs that differ in $\chi^2$ values. 
To obtain a range of acceptable fits, we use a cutoff value for $\chi^2$ that is described in \citet{Rob07}. 
We select the fit results which have $(\chi^2 - \chi^2_{best})/N_{data}$ $\leqq$ 3, where $\chi^2_{best}$ is the $\chi^2$ value of the best fit model and $N_{data}$ is the total number of data points used for the fit. 
Then, a median value of the selected results is adopted as a representative value and their standard deviation is adopted as uncertainty. 

In the \citet{Zha18} model, the evolution of the protostar and its surround structures in a self-consistent way based on the turbulent core accretion theory for massive star formation \citep{McK03}. 
The best models are selected based on $\chi^2$ values of the SED fits, as in the Robitaille model. 
We use best five models to estimate the final column density and uncertainty, i.e., their average value and standard deviation. 

In both models, the total visual extinction ($A_V$) from the protostar to the observer is derived from the best-fit SEDs. 
The value is doubled in order to compare with submillimeter data, which probe the total column density in the line of sight. 
To estimate $N_{\mathrm{H_2}}$ values from the derived $A_V$, we use a $N_{\mathrm{H_2}}$/$A_{V}$ conversion factor. 
\citet{Koo82} reported $N_{\mathrm{H}}$/$E(B-V)$ = 2.0 $\times$ 10$^{22}$ cm$^{-2}$ mag$^{-1}$ and \citet{Fit85} reported $N_{\mathrm{H}}$/$E(B-V)$ = 2.4 $\times$ 10$^{22}$ cm$^{-2}$ mag$^{-1}$ for the interstellar extinction in the LMC. 
Taking their average and adopting a slightly high $A_{V}$/$E(B-V)$ ratio of $\sim$4 for dense clouds \citep{Whi01b}, we obtain $N_{\mathrm{H_2}}$/$A_{V}$ = 2.8 $\times$ 10$^{21}$ cm$^{-2}$ mag$^{-1}$, where we assume that all the hydrogen atoms are in the form of H$_2$. 
The estimated $N_{\mathrm{H_2}}$ and $A_V$ are summarized in Table \ref{tab_N_H2}. 
For both SED models, consistent $N_{\mathrm{H_2}}$ values are obtained.

\subsubsection{$N_{\mathrm{H_2}}$ from the 9.7 $\mu$m silicate dust absorption} \label{sec_h2_tau97} 
The mid-infrared spectrum of ST16 shows a deep absorption due to the silicate dust at 9.7 $\mu$m (Fig. \ref{sed}). 
The peak optical depth of the 9.7 $\mu$m silicate dust absorption band ($\tau_{\mathrm{9.7}}$) is estimated to be 2.44 from the spectrum. 
The relationship between the visual extinction ($A_V$) and $\tau_{\mathrm{9.7}}$ reported for Galactic dense cores is 
\begin{equation}
A_V = \frac{\tau_{\mathrm{9.7}} - (0.12 \pm 0.05)}{0.21 \pm 0.02} \times 8.8 \label{Eq_Avtau} 
\end{equation}
according to \citet{Boo11} (assuming $A_V$/$A_K$ = 8.8). 
Applying this relationship to ST16, we obtain $A_V$ = 100 $\pm$ 10 mag. 
Because the present infrared absorption spectroscopy probes only the foreground component relative to the central protostar, the above $A_V$ value should be doubled to compare with submillimeter data, which probe the total column density in the line of sight. 
Therefore, the total visual extinction expected from the 9.7 $\mu$m silicate band is $A_{V}$ = 200 $\pm$ 20 mag for ST16. 
Using the LMC's $N_{\mathrm{H_2}}$/$A_V$ ratio of 2.8 $\times$ 10$^{21}$ cm$^{-2}$ mag$^{-1}$ described in Section \ref{sec_h2_sed}, we obtain $N_{\mathrm{H_2}}$ = (5.60 $\pm$ 0.56) $\times$ 10$^{23}$ cm$^{-2}$.

\subsubsection{Recommended H$_2$ column density, dust extinction, and gas mass} \label{sec_h2_final} 
The discussion in Section \ref{sec_h2_dust}--\ref{sec_h2_tau97} suggests that consistent $N_{\mathrm{H_2}}$ values are obtained by different methods; i.e., the SED fits by the \citet{Rob07}'s and the \citet{Zha18}'s model, and the $\tau_{\mathrm{9.7}}$--$A_V$ relation. 
In addition, the $N_{\mathrm{H_2}}$ estimates by the dust continuum with $T_{d}$ = 60 K result in consistent $N_{\mathrm{H_2}}$ values with the above methods. 
In this paper, we use $N_{\mathrm{H_2}}$ = (5.6 $\pm$ 0.6) $\times$ 10$^{23}$ cm$^{-2}$ as a representative value, which corresponds to the average of $N_{\mathrm{H_2}}$ derived by dust continuum (870 $\mu$m and 1200 $\mu$m with $T_{d}$ = 60 K), the SED model fits, and the 9.7 $\mu$m silicate dust absorption depth. 
This $N_{\mathrm{H_2}}$ corresponds to $A_V$ = 200 mag using the $N_{\mathrm{H_2}}$/$A_V$ factor described in Section \ref{sec_h2_sed}. 
Assuming the source diameter of 0.1 pc and the uniform spherical distribution of gas around a protostar, we estimate the average gas number density to be $n_{\mathrm{H_2}}$ = 3 $\times$ 10$^6$ cm$^{-3}$ and the total gas mass to be 100 M$_{\sun}$. 

Here we emphasize that the derived H$_2$ value corresponds to the total column density integrated over the whole line of sight, which includes various temperature components. 
Thus the assumed dust temperature ($T_{d}$ = 60 K) corresponds to the mass-weighted average temperature in the line of sight. 
Given the lower dust temperature compared with the gas temperature in the hot core region, the contribution from low-temperature component would not be negligible. 
The situation is the same for Galactic hot core sources compared in this work, whose N$_{\mathrm{H_2}}$ values are derived by using low-J CO isotopologue lines, and thus the low-temperature component would have non-negligible contribution. 
To selectively probe the total gas column density in the high-temperature hot core region, observations of high-J CO lines or H$_2$O lines will be important, which will be accessible by future far-infrared facilities. 

\subsection{Fractional abundances} \label{sec_x} 
Fractional abundances of molecules relative to H$_2$ are summarized in Table \ref{tab_X}, which are calculated by using the molecular column densities estimated in Section \ref{sec_n} and $N_{\mathrm{H_2}}$ estimated in Section \ref{sec_h2}. 
Abundances of HCN and HCO$^{+}$ are estimated from their isotopologues H$^{13}$CN and H$^{13}$CO$^{+}$, assuming $^{12}$C/$^{13}$C = 49 \citep{Wan09}. 

\begin{deluxetable}{ l c } 
\tablecaption{Estimated fractional abundances \label{tab_X}} 
\tablewidth{0pt} 
\tabletypesize{\small} 
\tablehead{
\colhead{Molecule}    & \colhead{$N$(X)/$N_{\mathrm{H_2}}$\tablenotemark{$\dag$}}
}
\startdata 
H$_2$CO                              &     (3.8 $\pm$ 0.6) $\times$ 10$^{-10}$             \\
CH$_3$OH\tablenotemark{a,b}   &    (4.8 $\pm$ 0.9) $\times$ 10$^{-9}$     \\
HCO$^+$\tablenotemark{c}     &    (1.2 $\pm$ 0.2) $\times$ 10$^{-9}$         \\
CCH                                    &     (2.3 $\pm$ 0.4) $\times$ 10$^{-10}$             \\
c-C$_3$H$_2$                       &    $<$4.3 $\times$ 10$^{-10}$                   \\
CN                                      &     (8.0 $\pm$ 1.8) $\times$ 10$^{-11}$             \\
HCN\tablenotemark{d}            &    (7.7 $\pm$ 2.6) $\times$ 10$^{-10}$      \\
NO                                       &    (7.3 $\pm$ 2.2) $\times$ 10$^{-9}$   \\
HNCO                                   &    (6.6 $\pm$ 2.2) $\times$ 10$^{-11}$   \\
CH$_3$CN\tablenotemark{a}   &    (4.1 $\pm$ 3.0) $\times$ 10$^{-11}$    \\
HC$_3$N                               &     $<$2.0 $\times$ 10$^{-11}$               \\
CS                                        &    (2.3 $\pm$ 0.4) $\times$ 10$^{-10}$    \\
H$_2$CS                              &     (3.4 $\pm$ 1.1) $\times$ 10$^{-11}$            \\
SO                                        &    (1.3 $\pm$ 0.2) $\times$ 10$^{-8}$    \\
SO$_2$\tablenotemark{a,b}    &    (1.1 $\pm$ 0.2) $\times$ 10$^{-8}$     \\
OCS\tablenotemark{a}            &    (1.2 $\pm$ 0.7) $\times$ 10$^{-9}$    \\  
SiO                                       &    (1.5 $\pm$ 0.7) $\times$ 10$^{-11}$   \\
HDO                                     &    $<$6.8 $\times$ 10$^{-10}$                  \\
C$_2$H$_5$OH                     &     $<$7.3 $\times$ 10$^{-10}$                  \\
C$_2$H$_5$CN                     &     $<$8.0 $\times$ 10$^{-10}$                 \\
CH$_3$OCH$_3$                  &     $<$4.5 $\times$ 10$^{-10}$                    \\
HCOOCH$_3$                       &     $<$1.2 $\times$ 10$^{-9}$                   \\
\textit{trans}-HCOOH              &     $<$1.3 $\times$ 10$^{-10}$                   \\
\enddata
\tablecomments{
$\dag$Assuming $N_{\mathrm{H_2}}$ = (5.6 $\pm$ 0.6) $\times$ 10$^{23}$ cm$^{-2}$. 
Molecular column densities are summarized in Table \ref{tab_N}. 
$^a$Based on the rotation analysis. 
$^b$Sum of all temperature components. 
$^c$Estimated from H$^{13}$CO$^+$ with $^{12}$C/$^{13}$C = 49. 
$^d$Estimated from H$^{13}$CN. 
} 
\end{deluxetable}

\section{Discussion} \label{sec_disc} 
\subsection{Hot molecular core associated with ST16} \label{sec_disc_hc} 
\begin{figure}[t]
\begin{center}
\includegraphics[width=9.0cm]{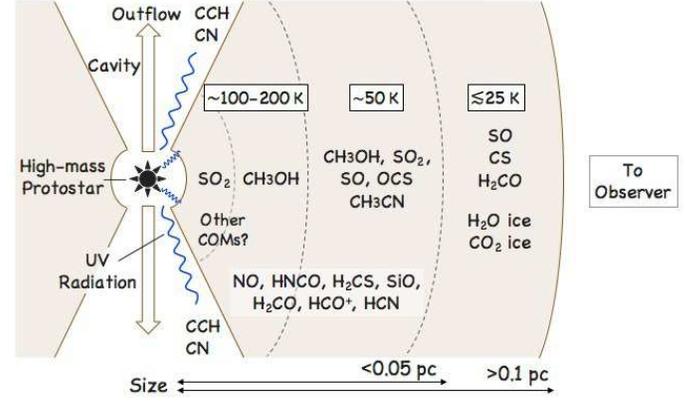}
\caption{
Schematic illustration of the temperature structure and the molecular distribution in the ST16 hot core. 
}
\label{structure}
\end{center}
\end{figure}

A variety of molecular emission lines, including high-excitation lines of typical hot core tracers such as CH$_3$OH and SO$_2$, are detected from the line of sight towards a high-mass YSO, ST16 (L = 3 $\times$ 10$^5$ L$_{\sun}$, Fig. \ref{sed}). 
The source is associated with high-density gas, as the H$_2$ gas density is estimated to be $n_{\mathrm{H_2}}$ = 3 $\times$ 10$^6$ cm$^{-3}$ based on the dust continuum data (Section \ref{sec_h2_final}). 
According to the rotation analyses of CH$_3$OH and SO$_2$ (Fig. \ref{rd}), the temperature of molecular gas is estimated to be higher than 100 K, which is sufficient to trigger the sublimation of ice mantles. 
The size of the hot gas emitting region is as compact as $\sim$0.1 pc, according to integrated intensity maps shown in Figures \ref{images1} and \ref{images2}. 
Note that the line of sight towards ST16 also contains compact and warm ($\sim$50--60 K) gas components as seen in the rotation diagrams of CH$_3$OH, SO$_2$, $^{34}$SO, OCS, and CH$_3$CN. 
In addition, the source is surrounded by relatively extended and cold ($\sim$25 K) gas components, as represented by the rotation diagram of SO at the off-center position. 
The nature of ST16, (i) the compact source size, (ii) the high gas temperature, (iii) the high density, (iv) the association with a high-mass YSO, (v) the presence of chemically-rich molecular gas, strongly suggests that the source is associated with a hot molecular core. 
The temperature structure and the molecular distribution in ST16 are illustrated in Figure \ref{structure} based on the present observational results. 

\subsection{Molecular abundances} \label{sec_disc_molab} 
\begin{figure}[tp]
\begin{center}
\includegraphics[width=9.0cm]{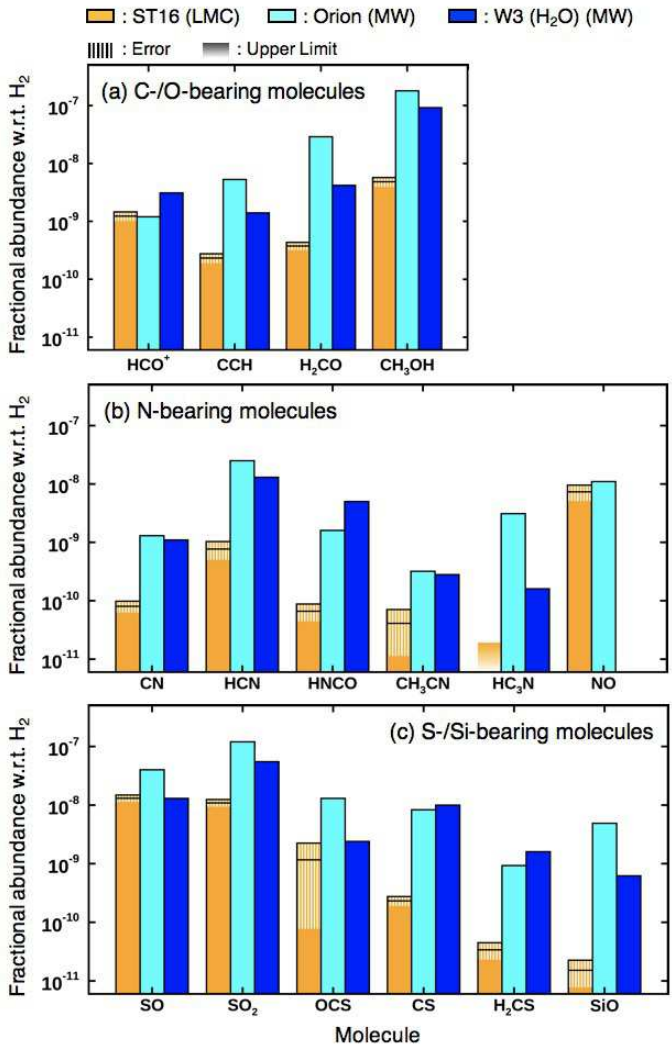}
\caption{
Comparison of molecular abundances between a LMC hot core (ST16: orange) and Galactic hot cores (Orion: cyan and W3 (H$_2$O): blue). 
Each panel shows (a) carbon- and oxygen-bearing molecules (HCO$^{+}$, CCH, H$_2$CO, and CH$_3$OH); (b) nitrogen-bearing molecules (CN, HCN, HNCO, CH$_3$CN, HC$_3$N, and NO); (c) sulfur- and silicon-bearing molecules (SO, SO$_2$, OCS, CS, H$_2$CS, and SiO). 
The area with thin vertical lines indicate the error bar. 
The bar with a color gradient indicate an upper limit. 
The plotted data are summarized in Table \ref{tab_X} for ST16, while those for Galactic hot cores are collected from the literature (see Section \ref{sec_disc_molab}). 
}
\label{histo_ab1}
\end{center}
\end{figure}

Figure \ref{histo_ab1} shows a comparison of molecular abundances between the ST16 hot core and Galactic hot cores (Orion and W3 (H$_2$O)). 
Abundances for Galactic hot cores are collected from \citet{Bla87, Tur91, Ziu91, Sut95, Schi97, Hel97}. 
Typically a factor of about two scatter in standard deviation is seen among molecular abundances of a number of Galactic hot cores \citep[e.g.,][]{Bis07,Ger14}. 
In general, fractional molecular abundances of ST16 are smaller than those of Galactic hot cores, but the degree of the abundance decrease varies depending on molecular species. 

Regarding carbon-bearing molecules shown in Figure \ref{histo_ab1}a, the H$_2$CO and CH$_3$OH abundances are low in the ST16 hot core by an order of magnitude or more, as compared with Galactic hot cores. 
CCH shows a lower abundance in ST16, but the difference is only a several factor relative to Galactic values. 
The HCO$^{+}$ abundance is comparable with those of Galactic sources. 

Nitrogen-bearing molecules are mostly less abundant than those in Galactic hot cores (Fig. \ref{histo_ab1}b). 
The degree of the abundance decrease is nearly a factor of five to ten. 
An exception is NO, whose abundance is comparable with Galactic values. 

For sulfur- and silicon-bearing molecules, the abundances of CS, H$_2$CS, and SiO are significantly less abundant in ST16 by more than an order of magnitude compared with Galactic hot cores (Fig. \ref{histo_ab1}c). 
The SO$_2$ abundance is lower by a factor of $\sim$4, while that of SO is not significantly different from Galactic values. 
The OCS abundance in ST16 may be comparable with or lower than Galactic values, but the abundance uncertainty is very large. 

\begin{figure*}[btp]
\begin{center}
\includegraphics[width=18.0cm]{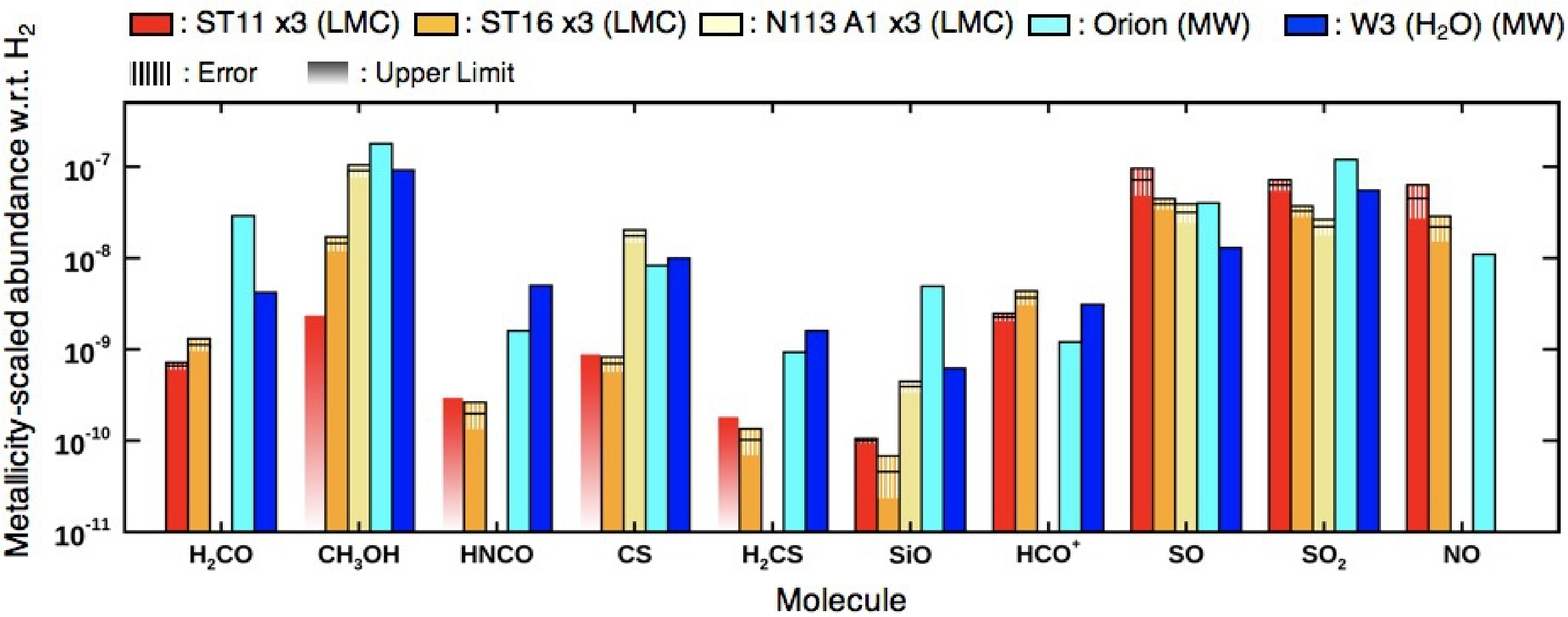}
\caption{
Comparison of metallicity-scaled molecular abundances between three LMC hot cores (ST11: red, ST16: orange, and N113 A1: light yellow) and Galactic hot cores (Orion: cyan) and W3 (H$_2$O): blue). 
Abundances of LMC hot cores are multiplied by three to correct for the metallicity difference relative to Galactic ones. 
The area with thin vertical lines indicate the error bar. 
The bar with a color gradient indicate an upper limit. 
The data for N113 A1 are adapted from \citet{Sew18}. 
See Section \ref{sec_disc_molab} for details. 
}
\label{histo_ab2}
\end{center}
\end{figure*}

Figure \ref{histo_ab2} compares metallicity-scaled fractional molecular abundances between LMC hot cores (ST11, ST16, N113 A1) and Galactic hot cores. 
The data for ST11 are adapted from \citet{ST16b} and those for N113 A1 are from \citet{Sew18}. 
The $N_{\mathrm{H_2}}$ value of N113 A1 is re-estimated using the same dust opacity data as in this work and the dust temperature assumption of 100 K; i.e. 5.3 $\times$ 10$^{23}$ cm$^{-2}$. 
Though it is not shown here, the other LMC hot core in \citet{Sew18}, N113 B3, shows a similar molecular abundance with N113 A1. 
In the figure, the abundances of LMC hot cores are multiplied by three to correct for the metallicity difference.  
Thus, the molecular abundance would roughly scale with the elemental abundances linearly, if the abundances are comparable between LMC and Galactic sources in the plot. 
If the metallicity-corrected molecular abundances of LMC sources are significantly higher or lower compared with Galactic counterparts, this would suggest that their overabundance or underabundance cannot be simply explained by the metallicity difference. 

It is commonly seen in two LMC hot cores (ST11 and ST16) that H$_2$CO, CH$_3$OH, HNCO, CS, H$_2$CS, and SiO are significantly less abundant, while HCO$^{+}$, SO, SO$_2$, and NO are comparable with or more abundant than Galactic hot cores, after corrected for the metallicity (Fig. \ref{histo_ab2}). 
The molecular abundances of the N113 A1 hot core in the LMC seem to be partly different from those in the other two LMC hot cores. 

The deficiency of organic molecules such as CH$_3$OH, H$_2$CO, and HNCO are previously reported for the ST11 hot core in the LMC \citep{ST16b}. 
A similar trend is seen in ST16, but the CH$_3$OH abundance in the N113 A1 hot core is almost comparable with Galactic values after corrected for the metallicity difference, as pointed out in \citet{Sew18}. 
In N113 A1 and B3, complex organic molecules larger than CH$_3$OH (i.e. CH$_3$OCH$_3$ and HCOOCH$_3$) are also detected \citep{Sew18}. 
This would suggest that organic molecules show a large abundance variation in a low-metallicity environment; ST11 and ST16 are organic-poor hot cores that are unique in the LMC and their low abundances of organic molecules cannot be explained by the decreased abundance of carbon and oxygen, while N113 A1 and B3 are organic-rich hot cores, whose COMs abundances roughly scale with the LMC's metallicity. 
It should be also noted that an infrared dark cloud that shows the CH$_3$OH abundance comparable with Galactic counterparts is detected in the SMC with ALMA \citep{ST18}. 
Although the source is not a hot core, this would be an alternative example indicating a large chemical diversity of organic molecules in low-metallicity environments. 

\citet{ST16, ST16b} suggest the warm ice chemistry hypothesis to interpret the low abundance of organic molecules in the LMC. 
The hypothesis argues that warm dust temperatures in the LMC inhibit the hydrogenation of CO in ice-forming dense clouds, which leads to the low abundances of organic molecules that are mainly formed on grain surfaces (CH$_3$OH, HNCO, and partially H$_2$CO). 
Therefore, the different chemical history during the ice formation stage could contribute to the differentiation of organic-poor and organic-rich hot cores in low-metallicity environments. 
Alternatively, the difference in the hot core's evolutionary stage may contribute to the observed chemical diversity, since high-temperature gas-phase chemistry can also decrease the CH$_3$OH abundance at a late stage \citep[e.g.,][]{NM04,Gar06,Vas13,Bal15}. 

The compact spatial distribution of CH$_3$OH lines in ST16 suggests its grain surface origin, as the emission is concentrated at the hot core position (Fig. \ref{images2}). 
On the other hand, the H$_2$CO emission is relatively extended around the hot core. 
This would suggest that H$_2$CO can also be formed efficiently in the gas phase, as in Galactic star-forming regions \citep[e.g.,][]{vdT00}. 

SO$_2$ is suggested to be a useful molecular tracer for the study of hot core chemistry at low metallicity, according to the ALMA observations of the ST11 hot core  \citep{ST16b}. 
This is because (i) SO$_2$ mainly originates from the hot core region as suggested from its compact distribution and high rotation temperature, 
(ii) the abundance of SO$_2$ simply scales with metallicity (the metallicity-scaling law), 
and (iii) SO$_2$ and its isotopologues show a large number of emission lines even in a limited frequency coverage. 
The same characteristics are observed in the present ST16 hot core. 
Metallicity-scaled abundances of SO$_2$ in LMC hot cores (ST11, ST16, N113 A1) are similar to each other and almost comparable with those of Galactic hot cores (Fig. \ref{histo_ab2}). 
This is remarkably in contrast to the abundance of a classical hot core tracer, CH$_3$OH, which shows a large variation in the low-metallicity environment of the LMC. 
The above characteristic behavior of SO$_2$ suggests that high-excitation SO$_2$ lines will be a useful tracer of metal-poor hot core chemistry. 

The metallicity-scaling law of SO$_2$ is not applicable to other sulfur-bearing molecules, as CS and H$_2$CS are significantly less abundant in organic-poor LMC hot cores. 
The metallicity-scaled abundances of SO in LMC hot cores are comparable with Galactic values, as in SO$_2$. 
However, the low rotation temperature and the extended spatial distribution of SO would suggest that it is widely distributed beyond the hot core region. 

NO may be an interesting molecule that is efficiently produced in low-metallicity environments. 
An overabundance of NO is reported for the ST11 hot core in the LMC \citep{ST16b}. 
Interestingly, a similar trend is observed in the present hot core. 
The metallicity-scaled abundance histogram shows that the NO abundances in LMC hot cores are higher than that in Orion, despite the low abundance of elemental nitrogen in the LMC (Fig. \ref{histo_ab2}). 
Note the plotted NO abundance of Orion is 1.1 $\times$ 10$^{-8}$, while the average and the standard deviation of NO abundances in six Galactic high-mass star-forming region is (8.2 $\pm$ 0.3) $\times$ 10$^{-9}$ \citep{Ziu91}. 
Currently, such a high abundance of nitrogen-bearing species is seen only in NO, as all of the other nitrogen-bearing molecules detected in ST16 are less abundant than those of Galactic hot cores (Fig. \ref{histo_ab1}b). 

The expected strength ratio of NO lines at 351.04352 and 351.05171 GHz in the LTE and optically thin case is 1.00 : 1.27, while the observed integrated intensity ratio is (1.00 $\pm$ 0.19) : (1.35 $\pm$ 0.10) based on the data in Table \ref{tab_lines_others}. 
Thus the observed NO lines are presumably optically thin. 

The distribution of NO is mainly concentrated at the hot core position, but slightly extended components are also seen at the east side of the hot core. 
The abundance of NO at the hot core position is (7.3 $\pm$ 2.2) $\times$ 10$^{-9}$ (see Table \ref{tab_X}). 
At the secondary peak on the east side, the abundance is estimated to be (7.5 $\pm$ 2.2) $\times$ 10$^{-9}$, where we assume $T_{\mathrm{rot}}$ = $T_{d}$ = 22 K based on the rotation analysis of SO, and with this temperature we have derived $N(\mathrm{NO})$ = (1.2 $\pm$ 0.2) $\times$ 10$^{15}$ cm$^{-2}$ and $N_{\mathrm{H_2}}$ = (1.6 $\pm$ 0.2) $\times$ 10$^{23}$ cm$^{-2}$. 
Similar abundances of NO between the hot core and the nearby peak may suggest a common chemical origin.

\begin{figure*}[tbph]
\begin{center}
\includegraphics[width=14.0cm]{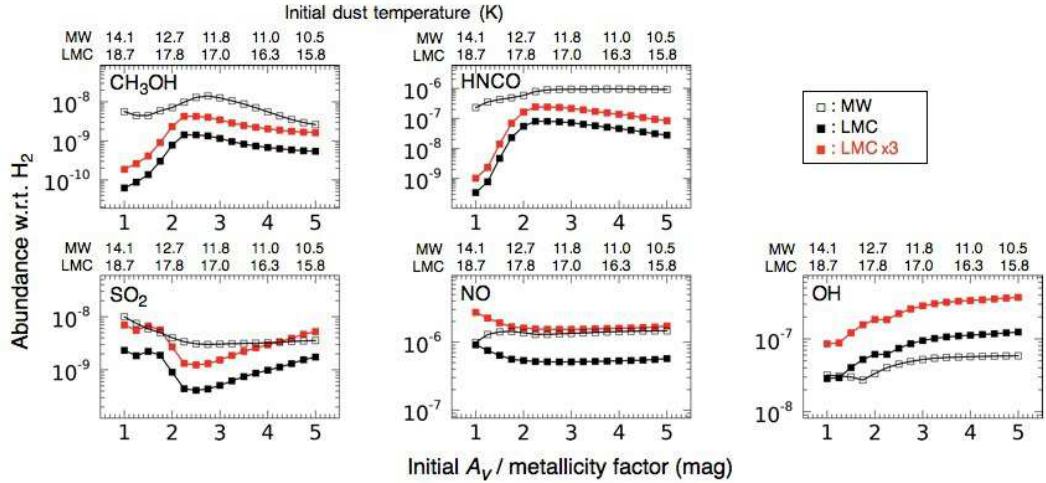}
\caption{
Simulated peak molecular abundances of CH$_3$OH, HNCO, SO$_2$, NO, and OH during the hot core stage plotted against the initial dust extinction at the prestellar stage. 
The corresponding dust temperature is also plotted at the upper axis in each panel. 
Note that, for LMC simulations, $A_V$ values are divided by three (metallicity factor) to mimic the low dust-to-gas ratio. 
$A_V$/(metallicity factor) = 1 mag corresponds to the gas column density of $N_{\mathrm{H_2}}$ = 2.8 $\times$ 10$^{21}$ cm$^{-2}$ using the $N_{\mathrm{H_2}}$/$A_V$ conversion factor in Section \ref{sec_h2_sed}. 
Results of hot core simulations for the Galactic case (open squares) and the LMC case (filled square) are plotted. 
The red filled squares represents the metallicity-scaled abundances for the LMC case, where the abundances are multiplied by three. 
}
\label{peak_vs_Av}
\end{center}
\end{figure*}

\subsection{Astrochemical Simulations} \label{sec_disc_theory} 
In this section we use astrochemical simulations to interpret the observed chemical characteristics of low-metallicity hot cores in the LMC. 
The simulations include a gas-grain chemical network coupled with a toy physical model, aiming at simulating the chemical evolution of a star-forming core up to the hot core stage. 
We here consider three different evolutionary stages (i.e., cold, warm-up, and post-warm-up), where physical conditions (density, temperature, and extinction) vary at each stage. 
The first stage corresponds to the quiescent cold cloud, the second stage mimics the collapsing and warming-up core, and the third stage corresponds to the high-temperature core before the formation of \ion{H}{2} region. 
The second and third stages correspond to the hot core. 
Two different setups for initial elemental abundances and gas-to-dust ratio are considered to simulate hot core chemistry at the LMC and Galactic conditions. 
Details of our astrochemical simulations are described in Section \ref{app_model} in Appendix. 

The chemical history during the ice forming stage is suggested to play an important role in subsequent hot core chemistry (Section \ref{sec_disc_molab}). 
Thus here we focus on the effect of initial physical conditions on the chemical compositions of a hot core. 
In the astrochemical simulations, we have varied the dust extinction parameter ($A_V$) at the first cold stage, and examined how the subsequent hot core compositions are affected. 
Here $A_V$ values are coupled with grain temperature using Eq. \ref{Eq_Hoc} in Appendix. 

Figure \ref{peak_vs_Av} shows the peak molecular abundances that are achieved during the hot core stage. 
The plotted peak abundances corresponds to the maximum achievable abundances of hot cores at each condition since hot core chemistry is time-dependent. 

It is seen from the figure that the maximum achievable abundances of CH$_3$OH gas in a hot core significantly decrease as the visual extinction of the first cold stage (ice-forming stage) decreases. 
The decrease of the abundance at the low $A_V$ regime is particularly enhanced in the LMC condition. 

The main reason behind this is because the abundance of solid CH$_3$OH is sensitive to the dust temperature. 
As discussed in the previous section, CH$_3$OH is mainly formed by grain surface reaction at the cold stage, and then released into the gas-phase at the hot core stage. 
The hydrogenation reaction of CO mainly controls the formation of CH$_3$OH on surfaces, but this reaction is inhibited when the grain surface temperature increases, because of the high volatility of atomic hydrogen. 
The effect is enhanced in the LMC case, because of the lower $A_V$ at the ice-forming stage, which leads to the higher dust temperature. 
A similar behavior is also seen in HNCO, suggesting the importance of hydrogenation for its production. 
The present simulations are consistent with the picture provided by the warm ice chemistry hypothesis \citep{ST16, ST16b} and previous astrochemical simulations dedicated to the LMC condition \citep{Ach15,Ach18,Pau18}. 

At the high $A_V$ regime, the peak CH$_3$OH abundances of the LMC case gradually approaches to the metallicity-corrected Galactic CH$_3$OH abundances. 
This could be because, in both LMC and Galactic cases, the grain surface is cold enough to trigger the CO hydrogenation, and the resultant CH$_3$OH abundances are regulated by the elemental abundances. 

The present astrochemical simulations suggest that a large chemical diversity of organic molecules seen in LMC hot cores is related to the different physical condition at the initial stage of star formation. 
Particular organic molecules such as CH$_3$OH  and HNCO decrease when a prestellar ice-forming cloud is less shielded, because of the inhibited surface hydrogenation reaction at increased dust temperature. 
This effect is particularly enhanced in a low-metallicity condition due to the low dust content in a star-forming core.

Molecular species that are mainly produced by high-temperature gas-phase chemistry show different behavior. 
SO$_2$ is one of those cases. 
It is suggested to be a key hot core tracer at low metallicity since the metallicity-scaling law can apply to its abundances (see Section \ref{sec_disc_molab}). 
Such a tendency is seen in the present simulation results. 
The peak abundances of SO$_2$ in a hot core, after corrected for the metallicity, are nearly comparable with Galactic cases (Fig. \ref{peak_vs_Av}). 
Also, SO$_2$ abundances are less affected by the initial $A_V$ at the ice-forming stage. 

A major molecular reservoir of sulfur in the cold stage is solid H$_2$S in our simulations. 
It is released into the gas phase at the hot core stage and experiences subsequent chemical syntheses into SO$_2$; 
i.e., H$_2$S $\xrightarrow{\mathrm{H}}$ SH $\xrightarrow{\mathrm{H}}$ S $\xrightarrow{\mathrm{OH/O_2}}$ SO $\xrightarrow{\mathrm{OH}}$ SO$_2$ \citep[e.g.,][]{Cha97,NM04}. 
This chemical sequence can reset the ice compositions that were accumulated at the cold stage and help initialize a major sulfur reservoir into atomic sulfur. 
Atomic sulfur is further synthesized into SO$_2$, but since it is a major product, the SO$_2$ abundance might be directly regulated by the elemental abundance of sulfur, which is roughly proportional to the metallicity. 
We speculate this \textit{reset} effect contributes to metallicity-scaled hot core compositions of particular molecular species.

NO in hot cores is also suggested to be mainly formed by high-temperature gas-phase chemistry, which is the neutral-neutral reaction between N and OH \citep[e.g.,][]{Her73,Pin90,NM04}. 
A major molecular reservoir of nitrogen in the cold stage is either solid N$_2$ or NH$_3$ in our simulations. 
Likewise SO$_2$, the parent species leading to the formation of NO would experience the chemical reset at the hot core stage. 
This could be the reason why NO abundances in a hot core are less affected by the physical condition of the initial ice-forming stage (Fig. \ref{peak_vs_Av}). 

Note that the simulated peak NO abundances in the LMC case become comparable with those of the Galactic case at the low $A_V$ regime. 
The behavior is consistent with the present observations, as LMC hot cores show NO abundances that are nearly comparable with Galactic hot cores despite the low nitrogen abundance. 
We speculate that the increased production of gaseous OH at the hot core stage may contribute to this, since peak OH abundances are higher in the LMC simulations as shown in Figure \ref{peak_vs_Av}. 
The efficient production of OH could be related to lower O/H or O/H$_2$ ratios at low metallicity. 
Alternatively, the increased production of solid NO at the prestellar stage may also contribute to the overproduction of NO in LMC hot cores. 
Gas-grain astrochemical simulations of a cold molecular cloud with a new set of atomic binding energies actually have reported the increase of solid NO according to the increased grain temperature \citep{STNN18}. 

More detailed and sophisticated astrochemical simulations of low-metallicity hot cores, involving the latest chemical data and various physicochemical mechanisms, are required for more quantitative interpretation of observed hot core compositions at different metallicity. 
Further simulations will be presented in a future paper.

\subsection{Infrared spectral characteristics of ST16} \label{sec_disc_ir} 
The observed hot core region corresponds to the infrared center of ST16. 
No emission line components (i.e. hydrogen recombination lines or fine-structure lines from ionized metals) are seen in the near- to mid-infrared spectrum of ST16 \citep{Sea09,ST16}. 
Despite the high bolometric luminosity, the source is still in an early evolutionary stage before the formation of a prominent \ion{H}{2} region. 
This would indicate that the central massive protostar has a low effective temperature ($<$10,000{\rm\:K}) and large radius ($>$100 $R_\odot$), which is theoretically predicted in the case of a high accretion rate with $>$$10^{-3}\:M_\odot{\rm\:yr^{-1}}$ \citep{Hos09,Tan18}. 

Abundances of solid molecules in the line-of-sight towards the infrared center of ST16 are summarized in Table \ref{tab_ice}. 
Elemental abundances of gas-phase oxygen and carbon in dense clouds in the LMC, after considering the depletion into dust grain material, are estimated to be 4.0 $\times$ 10$^{-5}$ for oxygen and 1.5 $\times$ 10$^{-5}$ for carbon (w.r.t. H$_2$), according to the LMC's low-metal abundance model presented in \citet{Ach15}. 
The total fractional abundance of elemental oxygen in solid H$_2$O and CO$_2$ (w.r.t. H$_2$) in ST16 is about 4.5 $\times$ 10$^{-6}$. 
This would suggest that a non-negligible fraction ($\sim$10 $\%$) of elemental oxygen still remain in ices in the direction of ST16. 
Because the temperature of the hot core region is high enough for the ice sublimation, these ices would exist in the ST16's cold outer envelope that is located at the foreground side to the observer. 
Note that, for more comprehensive estimates of gas/ice ratio, future observations of gas-phase H$_2$O, CO$_2$, and CO will be important.

\begin{deluxetable*}{ l c c c c}
\tablecaption{Ice abundances towards ST16 \label{tab_ice}} 
\tablewidth{0pt} 
\tabletypesize{\footnotesize} 
\tablehead{
\colhead{}                                      &  \colhead{H$_2$O ice}                       & \colhead{CO$_2$ ice}                      & \colhead{CO ice}                    & \colhead{CH$_3$OH ice}     
}
\startdata 
$N$(X) (10$^{17}$ cm$^{-2}$)      &  19.6 $\pm$ 3.2                                  &  2.7 $\pm$ 0.2                                  &  $<$2                                      &  $<$1.2            \\      
$N$(X)/$N_{\mathrm{H_2}}$        &  (3.5 $\pm$ 1.0) $\times$ 10$^{-6}$   & (4.8 $\pm$ 1.7) $\times$ 10$^{-7}$  &  $<$4 $\times$ 10$^{-7}$    &  $<$2.4 $\times$ 10$^{-7}$           \\
\enddata
\tablecomments{
Tabulated ice abundances are adapted from \citet{ST16}. We use $N_{\mathrm{H_2}}$ = (5.6 $\pm$ 0.6) $\times$ 10$^{23}$ cm$^{-2}$ based on this work. 
} 
\end{deluxetable*}

\subsection{Isotope abundances of sulfur} \label{sec_disc_isotop} 
Isotope abundances of $^{32}$S, $^{34}$S, and $^{33}$S, based on the present observations of SO, SO$_2$, CS, and their isotopologues, are summarized in Table \ref{tab_iso}. 

The $^{32}$S/$^{34}$S ratios for SO and SO$_2$ (23 and 15) in ST16 are well consistent with those estimated for the ST11 hot core and the N113 star-forming region in previous studies \citep[$\sim$15,][]{Wan09,ST16b}. 
This would suggests that both $^{32}$SO(3$_{3}$--2$_{3}$) and $^{32}$SO$_2$ lines are optically thin in the direction of ST16. 
The $^{32}$S/$^{34}$S ratio in the LMC sources is about a half compared with the solar neighborhood value \citep[$\sim$30,][]{Chi96b}, suggesting the overabundance of $^{34}$S in the LMC

The $^{32}$S/$^{33}$S ratios for SO and SO$_2$ (52 and 51) in ST16 are also, within the uncertainty, consistent with the previously-reported $^{32}$S/$^{33}$S ratio of 40 $\pm$ 17 estimated based on observations of the ST11 hot core \citep{ST16b}. 
The $^{32}$S/$^{33}$S ratio of SO and SO$_2$ in the LMC is significantly lower than a typical solar neighborhood value by a factor of 3--4. 
As well as $^{34}$S, $^{33}$S is also overabundant in the LMC. 

Sulfur is an $\alpha$ element and massive stars mainly contribute to its nucleosynthesis. 
A theoretical model on the galactic chemical enrichment predicts the increasing trend of $^{32}$S/$^{33, 34}$S ratios according to the decreasing metallicity, because minor isotopes are synthesized from the seed of major isotopes as secondary elements in core-collapse supernovae, and thus more minor isotopes are produced at higher metallicity \citep{Kob11}. 
The trend is consistent with the observations of sulfur isotopes in our Galaxy, which report an increase of the $^{32}$S/$^{34}$S ratio from the Galactic inner part toward the solar neighborhood \citep[see Fig.3 in][]{Chi96b}. 

The sulfur isotope ratios in the LMC, however, significantly deviate from this trend. 
The observed $^{32}$S/$^{34}$S and $^{32}$S/$^{33}$S ratios in ST16 are lower than those of the model prediction at a half solar metallicity by a factor of two and four, respectively \citep[see Table. 3 in][]{Kob11}. 
The LMC's sulfur isotope ratio also deviates from the above-mentioned increasing trend of the $^{32}$S/$^{34}$S ratio from the Galactic inner part to the solar neighborhood. 
Interestingly, low $^{32}$S/$^{34}$S ratios are observed in millimeter molecular absorption line systems at the redshift of 0.68 and 0.89 \citep[$^{32}$S/$^{34}$S $\sim$9 for CS/C$^{34}$S and H$_2$S/H$_2$$^{34}$S,][]{Wal16}. 
The reason of the characteristic isotope abundance ratios of sulfur in the LMC is still remain unexplained.

\begin{deluxetable*}{ l c c c c c c c}
\tablecaption{Isotope abundances of sulfur \label{tab_iso}} 
\tablewidth{0pt} 
\tabletypesize{\footnotesize} 
\tablehead{
\colhead{ }  & \multicolumn{4}{c}{ST16\tablenotemark{a}}                                       & \colhead{ST11\tablenotemark{b}}  & \colhead{N113\tablenotemark{c}} & \colhead{Solar neighborhood\tablenotemark{d}}  \\
\cline{2-5}
\colhead{}                    &  \colhead{SO}   & \colhead{SO$_2$}  & \colhead{CS}  & \colhead{Weighted Mean\tablenotemark{e}}   & \colhead{SO$_2$}    & \colhead{CS}                                 &  \colhead{CS}  
}
\startdata 
$^{32}$S/$^{34}$S      &  23 $\pm$ 8       &  15 $\pm$ 3           &  19 $\pm$ 3    &  17 $\pm$ 2   &  14 $\pm$ 3                                     &  $\sim$15                                      & $\sim$30  \\      
$^{32}$S/$^{33}$S      &  52 $\pm$ 5      &  51 $\pm$ 15         &  72 $\pm$ 18   &  53 $\pm$ 5   &  40 $\pm$ 17                                   &  $<$100                                        & $\sim$180  \\      
$^{34}$S/$^{33}$S      &  2 $\pm$ 1        &  3 $\pm$ 1              &  4 $\pm$ 1      &  3 $\pm$ 1    &  3 $\pm$ 2                                       &  $<$7                                            & $\sim$6  \\   
\enddata
\tablecomments{
$^a$This work; 
$^b$ALMA observations towards a LMC hot core, ST11 \citep{ST16b}; 
$^c$Single-dish observations towards a LMC star-forming region, N113 \citep{Wan09}; 
$^d$\citet{Chi96b}; 
$^e$The weight is the inverse of the squared error value. 
} 
\end{deluxetable*}
%

\subsection{Rotating protostellar envelope traced by $^{34}$SO and SO$_2$} \label{sec_disc_rotation} 
\begin{figure*}[btp]
\begin{center}
\includegraphics[width=18.0cm]{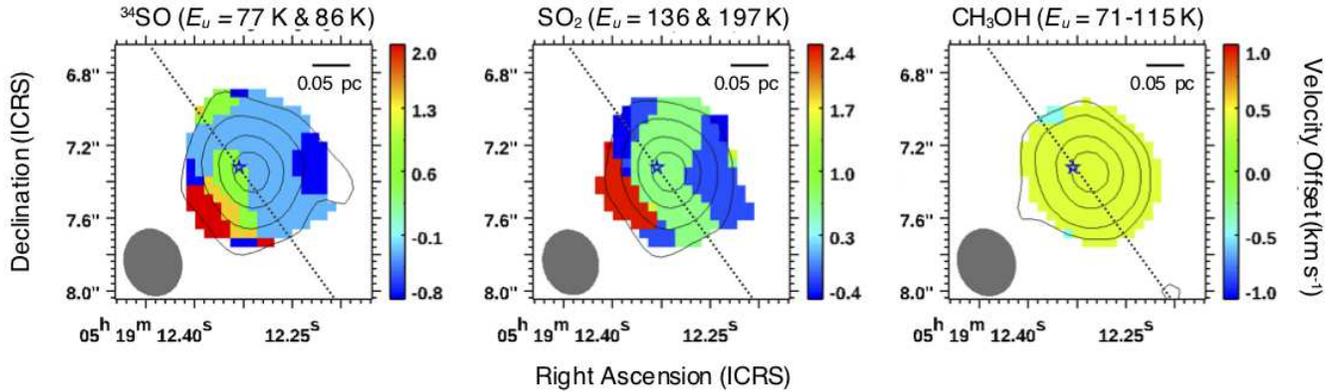}
\caption{
Velocity maps for $^{34}$SO, SO$_2$, and CH$_3$OH. 
The color scale indicates the offset velocity relative to the systemic velocity of 264.5 km s$^{-1}$. 
The contour represents the integrated intensity, where the level is 6 $\%$, 20 $\%$, 50 $\%$, and 80 $\%$ of the peak value for $^{34}$SO and SO$_2$, while 10 $\%$, 20 $\%$, 50 $\%$, and 80 $\%$ for CH$_3$OH. 
The dotted line represents a possible outflow axis inferred from CCH and CN distribution (see Section \ref{sec_disc_CCH_CN}). 
The maps are constructed by averaging the following Band 7 lines: $^{34}$SO($N_J$ = 8$_{8}$--7$_{7}$ and 8$_{9}$--7$_{8}$), SO$_2$(18$_{4,14}$--18$_{3,15}$ and 14$_{4,10}$--14$_{3,11}$), CH$_3$OH(7$_{0}$ E--6$_{0}$ E, 7$_{-1}$ E--6$_{-1}$ E, 7$_{2}$ A$^-$--6$_{2}$ A$^-$, 7$_{3}$ A$^+$--6$_{3}$ A$^+$, and 7$_{-2}$ E--6$_{-2}$ E). 
The blue open star represents the position of a high-mass YSO. 
The synthesized beam size is shown by the gray filled ellipse. 
See Section \ref{sec_disc_rotation} for details. 
}
\label{mom1b}
\end{center}
\end{figure*}

A sign of the rotating protostellar envelope is seen in the velocity maps of $^{34}$SO and SO$_2$ (Figure \ref{mom1b}). 
The maps are constructed based on the original Band 7 images without the beam restoration, where the beam size corresponds to 0.090 pc $\times$ 0.076 pc at the LMC. 
As shown in the figure, the east side of the core is red-shifted and the west side is blue-shifted, and the velocity separation is about 2--3 km s$^{-1}$. 
The direction of the velocity separation is nearly perpendicular to the outflow axis, which is directed from north-east to south-west (see Section \ref{sec_disc_CCH_CN}). 
This would support the idea that $^{34}$SO and SO$_2$ are tracing the envelope rotation, rather than the outflow motion. 
Similar rotation motions are observed in Galactic high-mass protostellar objects \citep[e.g.,][]{Beu07b,Fur10,Bel10,Bel11,Zha19}. 
The present result is the first detection of a rotating protostellar envelope outside our Galaxy. 

SO does not show the rotation motion in the present data. 
This may be due to the contamination of a foreground cold component, because the strong SO lines (6$_{6}$--5$_{5}$ and 8$_{7}$--7$_{6}$) are moderately optically thick (see Section \ref{sec_rd}), while the optically thin SO(3$_{3}$--2$_{3}$) is too weak to analyze the velocity structure. 

Figure \ref{mom1b} also shows the velocity map for high-excitation CH$_3$OH lines, but the rotating structure is not seen here, though these CH$_3$OH  trace a warm and dense region close to the protostar. 
This would indicate the different spatial distribution of CH$_3$OH and $^{34}$SO/SO$_2$ within the 0.1 pc scale region. 

\begin{figure}[t]
\begin{center}
\includegraphics[width=9.5cm]{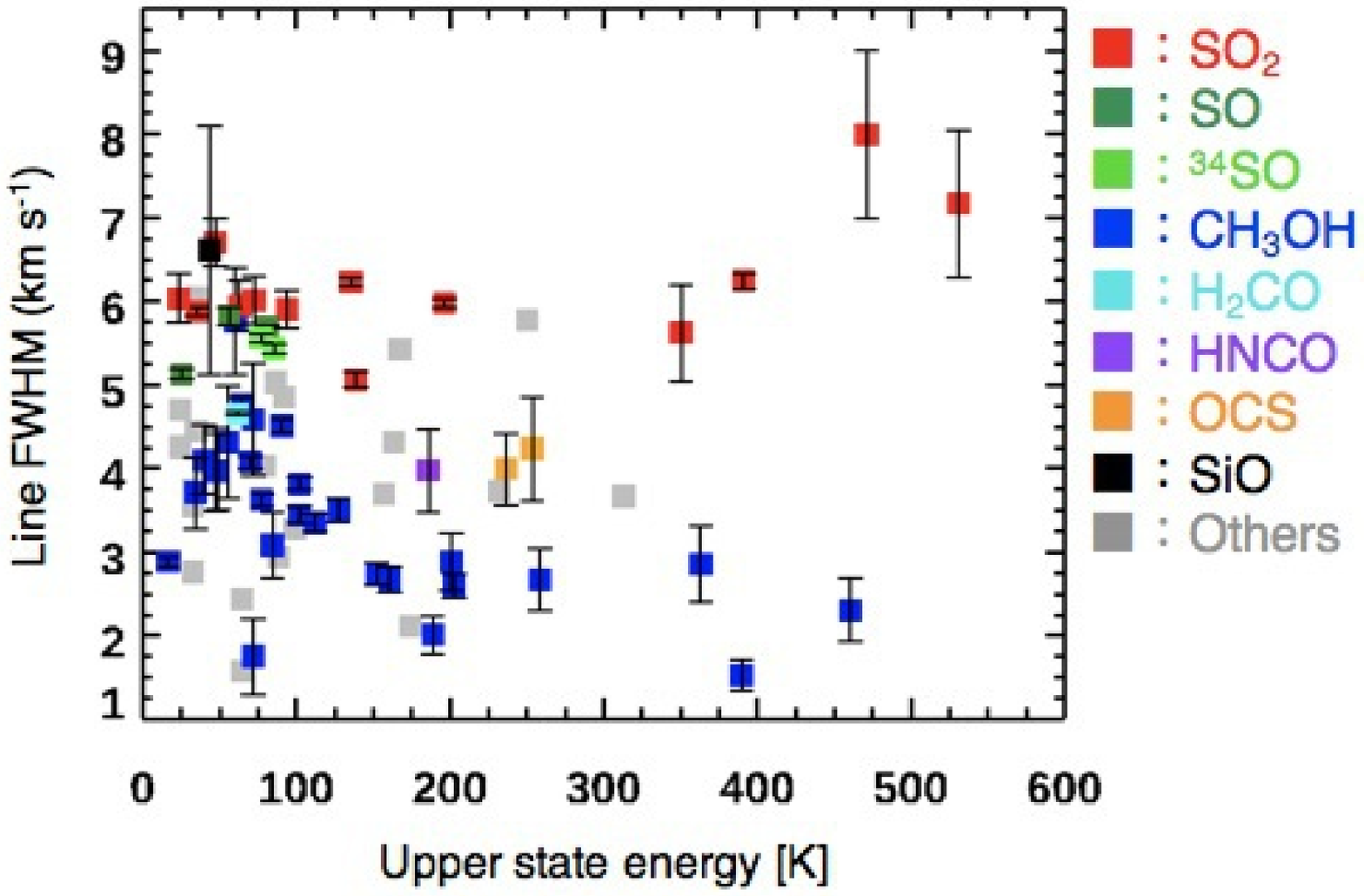}
\caption{
Line FWHMs vs. upper state energies. 
Blended lines and low-S/N lines are excluded. 
Molecular names are indicated by colors (SO$_2$: red, SO: dark green, $^{34}$SO: light green, CH$_3$OH: blue, H$_2$CO: cyan, HNCO: purple, OCS: orange, SiO: black, others: gray). 
}
\label{fwhm_Eu}
\end{center}
\end{figure}

The different distributions of the sulfur-bearing species and CH$_3$OH are also supported by their different line widths. 
Figure \ref{fwhm_Eu} compares the measured line FWHMs of the selected molecular species with their upper state energies. 
Blended lines and low-S/N lines are excluded here. 
The figure shows that SO$_2$, SO, and $^{34}$SO have relatively large line widths ($\gtrsim$6 km s$^{-1}$). 
A well-known shock tracer, SiO, also shows a broader line width. 
On the other, CH$_3$OH lines show relatively narrow line widths ($\lesssim$5 km s$^{-1}$ for $E_{u}$ $<$150 K and $\lesssim$3 km s$^{-1}$ for $E_{u}$ $>$150 K). 
Other molecules, H$_2$CO, HNCO, and OCS, which possibly trace a compact hot core region, show intermediate line widths between SO$_2$ and CH$_3$OH. 
The relatively large line widths of SO and SO$_2$ would indicate that they arise from a more turbulent region compared with other molecular species. 
Given the similar line width of SiO and SO/SO$_2$, such a turbulent region may be related to the shock. 

We finally note that an infall motion is not seen in the present data, because we do not cover a fully optically thick molecular tracer in this work. 
Future higher-spatial-resolution multiline observations are required to further understand the dynamics of molecular gas associated with a low-metallicity massive protostellar envelope.

\subsection{Outflow cavity structures traced by CCH and CN} \label{sec_disc_CCH_CN} 
Spatial distributions of molecular radicals, CCH and CN, are similar to each other. 
In the LTE and optically thin case, the expected intensity ratio of CCH lines at 349.33771 and 349.39928 GHz is 1.28 : 1.00, which is consistent with the observed integrated intensity ratio of (1.25 $\pm$ 0.04) : (1.00 $\pm$ 0.05) based on the data in Table \ref{tab_lines_others}, suggesting that the lines are optically thin. 
Similarly, for the same assumption, the expected ratio of CN lines at 340.03155, 340.03541, and 340.24777 GHz is 1.00 : 1.00 : 3.05, while the observed integrated intensity ratio is (1.00 $\pm$ 0.10) : (1.11 $\pm$ 0.11) : (2.75 $\pm$ 0.08), suggesting that they are nearly optically thin. 

Obviously the CCH and CN distributions are not centered at the hot core position. 
They show emission peaks at the north-east and south-west direction from the hot core. 
Their distribution seems to trace bipolar outflow structures, originating from the protostar associated with the hot core. 
A well-collimated structure is seen particularly in the north direction. 
A width of the collimated structure is $\sim$0.1 pc, almost the same as the beam size. 
Given the early evolutionary stage of ST16, it is likely that molecular outflows are associated with the source. 

CCH and CN emission are known to be bright in photodissociation regions (PDRs) and they are suggested to be a sensitive tracer of UV-irradiated regions \citep[e.g.,][]{Fue93, Jan95, Rod98, Pet17}. 
The present characteristic distributions of CCH and CN presumably trace the PDR-like outflow cavity structure, which are irradiated by the UV light. 
According to the present dust continuum data, the visual extinction from the outer edge of the dust clump to the CCH/CN emission region is at least larger than 20 mag. 
Thus a possible UV source for the photochemistry is a high-mass protostar located at the hot core position, rather than the external radiation field. 
Such a UV-irradiated outflow cavity is actually observed in Galactic star-forming regions via CCH emission \citep{Zha18b}. 
The presence of high-velocity outflow gas needs to be tested by future high-spatial-resolution observations of strong and optically thick outflow tracers such as CO.

\section{Summary} \label{sec_sum} 
We present the results of 0$\farcs$40 (0.1 pc) scale submillimeter observations towards a high-mass YSO (ST16, L = 3 $\times$ 10$^5$ L$_{\sun}$) in the LMC with ALMA. 
As a result, a new hot molecular core is discovered in the LMC. 
The following conclusions are obtained in this work: 

\begin{enumerate}
%
\item 
Molecular emission lines of CH$_3$OH, H$_2$CO, CCH, H$^{13}$CO$^{+}$, CS, C$^{34}$S, C$^{33}$S, SO, $^{34}$SO, $^{33}$SO, SO$_2$, $^{34}$SO$_2$, $^{33}$SO$_2$, OCS, H$_2$CS, CN, NO, HNCO, H$^{13}$CN, CH$_3$CN, and SiO are detected from the compact region ($\sim$0.1 pc) associated with a high-mass YSO. 
In total we have detected 90 transitions, out of which, 30 lines are due to CH$_3$OH, and 27 lines are due to SO$_2$ and its isotopologues. 
Complex organic molecules larger than CH$_3$OH are not detected. 
%

\item 
Rotation analyses show that ST16 is associated with hot molecular gas ($T_{\mathrm{rot}}$ $>$ 100 K) as traced by CH$_3$OH and SO$_2$. 
The line of sight also contains warm ($\sim$50--60 K) gas components traced by CH$_3$OH, SO$_2$, $^{34}$SO, OCS, and CH$_3$CN, in addition to extended and cold ($\sim$25 K) gas traced by SO. 
%

\item 
The total gas column density towards ST16 is estimated by using several different methods (continuum analysis, SED analysis, and dust absorption band analysis). 
The estimated H$_2$ column density is $N_{\mathrm{H_2}}$ = 5.6 $\times$ 10$^{23}$ cm$^{-2}$ (corresponds to $A_V$ = 200 mag). 
The average gas number density is estimated to be $n_{\mathrm{H_2}}$ = 3 $\times$ 10$^6$ cm$^{-3}$. 
%

\item 
The nature of ST16, the compact source size, the high gas temperature, the high density, association with a high-mass YSO, and the presence of chemically-rich molecular gas, strongly suggest that the source is associated with a hot molecular core. 
%

\item 
Organic molecules show a large abundance variation in low-metallicity hot cores. 
There are currently two organic-poor hot cores \citep[ST16 in this work and ST11 in][]{ST16b} and two organic-rich hot cores \citep[N113 A1 and B3 in][]{Sew18} in the LMC. 
The different chemical history during the ice formation stage could contribute to the differentiation of organic-poor and organic-rich hot cores. 
%

\item 
High-excitation SO$_2$ lines will be a useful tracer of low-metallicity hot core chemistry. 
This is because (i) SO$_2$ mainly originates from a hot core region, (ii) it is commonly seen in LMC hot cores, and (iii) its abundances in LMC hot cores roughly scale with the LMC's metallicity. 
This is remarkably in contrast to abundances of a classical hot core tracer, CH$_3$OH, which shows a large abundance variation in low-metallicity hot cores. 
CS and H$_2$CS are significantly less abundant in organic-poor hot cores. 
%

\item 
Nitrogen-bearing molecules in ST16 are generally less abundant than those in Galactic hot cores. 
An exception is NO, whose abundance is comparable with Galactic values, despite the low elemental abundance. 
An overabundance of NO is also reported in the other LMC hot core (ST11). 
%

\item 
Isotope abundance ratios of $^{32}$S, $^{33}$S, and $^{34}$S in the ST16 hot core are presented. 
Based on SO, SO$_2$, and their isotopologues, we obtain $^{32}$S/$^{34}$S $\sim$15 and $^{32}$S/$^{33}$S $\sim$40, which are lower than solar neighborhood values by a factor of 2 and 4.5, respectively. 
Both $^{34}$S and $^{33}$S are overabundant in the LMC. 
%

\item 
A rotating protostellar envelope is for the first time detected outside our Galaxy via SO$_2$ and $^{34}$SO lines. 

\item 
CCH and CN show clearly different spatial distributions compared to other molecular lines. 
They seem to trace PDR-like cavity regions created by protostellar outflows. 
%

\item 
Our astrochemical simulations for a low-metallicity hot core suggest that a large chemical diversity of organic molecules (e.g., CH$_3$OH) seen in LMC hot cores is related to the different physical condition at the initial stage of star formation. 
Particular molecular species that are mainly produced by high-temperature gas-phase chemistry in a hot core (e.g., SO$_2$) likely to show a metallicity-scaled molecular abundance. 

\end{enumerate}

\acknowledgments 
This paper makes use of the following ALMA data: ADS/JAO.ALMA$\#$2016.1.00394.S and $\#$2018.1.01366.S. 
ALMA is a partnership of ESO (representing its member states), NSF (USA) and NINS (Japan), together with NRC (Canada) and NSC and ASIAA (Taiwan) and KASI (Republic of Korea), in cooperation with the Republic of Chile. 
The Joint ALMA Observatory is operated by ESO, AUI/NRAO and NAOJ. 
This work has made extensive use of the Cologne Database for Molecular Spectroscopy and the molecular database of the Jet Propulsion Laboratory. 
We use data obtained by IRSF/SIRIUS, VLT/ISAAC, \textit{AKARI}, \textit{Spitzer}, and \textit{Herschel}. 
We are grateful to all the members who contributed to these projects. 
T.S. is supported by a Grant-in-Aid for Scientific Research on Innovative Areas (19H05067) and Leading Initiative for Excellent Young Researchers, MEXT, Japan. 
A.D acknowledges the ISRO RESPOND program (Grant No. ISRO/RES/2/402/16-17) and Grant-In-Aid from the Higher Education Department of the Government of West Bengal. 
K.E.I.T acknowledges support from NAOJ ALMA Scientific Research Grant Number 2017-05A and JSPS KAKENHI Grant Number JP19K14760. 
Y.N. is supported by NAOJ ALMA Scientific Research Grant Number 2017-06B and JSPS KAKENHI Grant Number JP18K13577. 
Finally, we would like to thank an anonymous referee for careful reading and useful comments.


\clearpage

\appendix

\restartappendixnumbering

\section{Measured line parameters}
\startlongtable
\begin{deluxetable*}{ l c c c c c c c c c }
\tablecaption{Line Parameters \label{tab_lines_others}}
\tablewidth{0pt}
\tabletypesize{\scriptsize} 
\tablehead{
\colhead{Molecule}   & \colhead{Transition}                               &       \colhead{$E_{u}$} &       \colhead{Frequency} &        \colhead{$T_{\mathrm{b}}$} &     \colhead{$\Delta$$V$} &     \colhead{$\int T_{\mathrm{b}} dV$} &       \colhead{$V_{\mathrm{LSR}}$} &        \colhead{RMS} &       \colhead{Note} \\
\colhead{ }          & \colhead{ }                                        &        \colhead{(K)} &           \colhead{(GHz)} &             \colhead{(K)} &          \colhead{(km s$^{-1}$)} &             \colhead{(K km s$^{-1}$)} &          \colhead{(km s$^{-1}$)} &        \colhead{(K)} &           \colhead{}
}
\startdata 
 CCH                                     &  N = 4--3, J = $\frac{9}{2}$--$\frac{7}{2}$, F = 5--4                                                                    &   42       & 349.33771       &   1.43 $\pm$   0.02  &    4.0          & 6.14 $\pm$ 0.22      & 263.9           & 0.06 &    (1) \\
 CCH                                     &  N = 4--3, J = $\frac{7}{2}$--$\frac{5}{2}$, F = 4--3                                                                    &   42       & 349.39928       &   1.15 $\pm$   0.02  &    4.0          & 4.93 $\pm$ 0.22      & 264.0           & 0.06 &    (2) \\
 H$_2$CO                                 &  5$_{1,5}$--4$_{1,4}$                                                                                                    &   62       & 351.76864       &   6.63 $\pm$   0.03  &    4.7          & 32.83 $\pm$ 0.31     & 264.5           & 0.06 &    \nodata \\
 H$^{13}$CO$^+$                          &  3--2                                                                                                                    &   25       & 260.25534       &   3.21 $\pm$   0.14  &    4.7          & 16.00 $\pm$ 1.50     & 264.7           & 0.29 &    \nodata \\
 CS                                         &  5--4                                                                                                                    &   35       & 244.93556       &   8.55 $\pm$   0.16  &    4.4          & 40.43 $\pm$ 1.59     & 264.4           & 0.30 &    \nodata \\
 C$^{34}$S                               &  7--6                                                                                                                    &   65       & 337.39646       &   0.88 $\pm$   0.03  &    2.4          & 2.28 $\pm$ 0.14      & 264.3           & 0.06 &    \nodata \\
 C$^{33}$S                               &  5--4                                                                                                                    &   35       & 242.91361       &             $<$0.60  &    \nodata      &             $<$2.6   &    \nodata      & 0.30 &    \nodata \\
 C$^{33}$S                               &  7--6                                                                                                                    &   65       & 340.05257       &   0.36 $\pm$   0.03  &    $<$2          & 0.60 $\pm$ 0.10      & 264.5           & 0.06 &    \nodata \\
 SO                                      &  $N_J$ = 6$_{6}$--5$_{5}$                                                                                                        &   56       & 258.25583       &   9.73 $\pm$   0.16  &    5.8          & 60.26 $\pm$ 2.16     & 264.8           & 0.29 &    \nodata \\
 SO                                      &  $N_J$ = 3$_{3}$--2$_{3}$                                                                                                        &   26       & 339.34146       &   2.90 $\pm$   0.03  &    5.1          & 15.79 $\pm$ 0.37     & 265.0           & 0.06 &    \nodata \\
 SO                                      &  $N_J$ = 8$_{7}$--7$_{6}$                                                                                                        &   81       & 340.71416       &  10.84 $\pm$   0.03  &    5.7          & 65.66 $\pm$ 0.42     & 264.7           & 0.06 &    \nodata \\
 $^{34}$SO                               &  $N_J$ = 8$_{8}$--7$_{7}$                                                                                                        &   86       & 337.58015       &   2.90 $\pm$   0.03  &    5.4          & 16.76 $\pm$ 0.32     & 264.9           & 0.06 &    \nodata \\
 $^{34}$SO                               &  $N_J$ = 3$_{3}$--2$_{3}$                                                                                                        &   25       & 337.89225       &   0.20 $\pm$   0.02  &    3.1          & 0.67 $\pm$ 0.17      & 265.2           & 0.06 &    \nodata \\
 $^{34}$SO                               & $N_J$ =  8$_{9}$--7$_{8}$                                                                                                        &   77       & 339.85727       &   3.19 $\pm$   0.02  &    5.5          & 18.82 $\pm$ 0.31     & 265.0           & 0.06 &    \nodata \\
 $^{33}$SO                               &  $N_J$ = 6$_{7}$--5$_{6}$                                                                                                    &   47       & 259.28403       &   1.48 $\pm$   0.58  &   \nodata          & 6.14 $\pm$ 0.76      & \nodata         & 0.29 &    (3) (4) \\
 $^{33}$SO                               &  $N_J$ = 8$_{8}$--7$_{7}$                                                                                                        &   87       & 340.83964       &   1.43 $\pm$   0.03  &    5.0          & 7.66 $\pm$ 0.37      & 266.0           & 0.06 &    (5) \\
 H$_2$CS                                 &  7$_{1,6}$--6$_{1,5}$                                                                                                    &   60       & 244.04850       &             $<$0.60  &    \nodata      &             $<$2.6   &    \nodata      & 0.30 &    \nodata \\
 H$_2$CS                                 &  10$_{1,10}$--9$_{1,9}$                                                                                                  &  102       & 338.08319       &   0.25 $\pm$   0.02  &    3.4          & 0.92 $\pm$ 0.19      & 264.7           & 0.06 &    \nodata \\
 OCS                                     &  20--19                                                                                                                  &  123       & 243.21804       &   0.89 $\pm$   0.16  &    5.7          & 5.40 $\pm$ 1.98      & 265.3           & 0.30 &    $\dag$ \\
 OCS                                     &  28--27                                                                                                                  &  237       & 340.44927       &   0.27 $\pm$   0.02  &    4.0          & 1.13 $\pm$ 0.22      & 263.1           & 0.06 &    \nodata \\
 OCS                                     &  29--28                                                                                                                  &  254       & 352.59957       &   0.22 $\pm$   0.02  &    4.2          & 1.00 $\pm$ 0.23      & 264.4           & 0.06 &    \nodata \\
 CN                                      &  N = 3--2, J = $\frac{5}{2}$--$\frac{3}{2}$, F = $\frac{7}{2}$--$\frac{5}{2}$                                            &   33       & 340.03155       &   0.96 $\pm$   0.03  &    2.8          & 2.82 $\pm$ 0.28      & 265.0           & 0.06 &    (6) \\
 CN                                      &  N = 3--2, J = $\frac{5}{2}$--$\frac{3}{2}$, F = $\frac{5}{2}$--$\frac{3}{2}$                                            &   33       & 340.03541       &   0.83 $\pm$   0.02  &    3.5          & 3.12 $\pm$ 0.32      & 264.8           & 0.06 &    (7) \\
 CN                                      &  N = 3--2, J = $\frac{7}{2}$--$\frac{5}{2}$, F = $\frac{9}{2}$--$\frac{7}{2}$                                            &   33       & 340.24777       &   1.95 $\pm$   0.03  &    3.7          & 7.75 $\pm$ 0.22      & 264.4           & 0.06 &    (8) \\
 H$^{13}$CN                              &  3--2                                                                                                                    &   25       & 259.01180       &   1.27 $\pm$   0.13  &    4.2          & 5.76 $\pm$ 1.31      & 265.0           & 0.29 &    \nodata \\
 NO                                      &  J = $\frac{7}{2}$--$\frac{5}{2}$, $\Omega$ = $\frac{1}{2}$, F = $\frac{9}{2}$$^-$--$\frac{7}{2}$$^+$                    &   36       & 350.68949       &             $<$0.90  &    \nodata      &             $<$3.8   &    \nodata      & 0.06 &    (9) (10) \\
 NO                                      &  J = $\frac{7}{2}$--$\frac{5}{2}$, $\Omega$ = $\frac{1}{2}$, F = $\frac{5}{2}$$^-$--$\frac{3}{2}$$^+$                    &   36       & 350.69477       &             $<$0.70  &    \nodata      &             $<$3.0   &    \nodata      & 0.06 &    (9) \\
 NO                                      &  J = $\frac{7}{2}$--$\frac{5}{2}$, $\Omega$ = $\frac{1}{2}$, F = $\frac{9}{2}$$^+$--$\frac{7}{2}$$^-$                    &   36       & 351.04352       &   0.33 $\pm$   0.02  &    6.0          & 2.12 $\pm$ 0.40      & 264.8           & 0.06 &   \\
 NO                                      &  J = $\frac{7}{2}$--$\frac{5}{2}$, $\Omega$ = $\frac{1}{2}$, F = $\frac{7}{2}$$^+$--$\frac{5}{2}$$^-$                    &   36       & 351.05171       &   0.82 $\pm$   0.02  &    3.3          & 2.87 $\pm$ 0.21      & 264.6           & 0.06 &    (11) \\
 HNCO                                    &  11$_{0,11}$--10$_{0,10}$                                                                                                &   70       & 241.77403       &             $<$1.00  &    \nodata      &             $<$4.3   &    \nodata      & 0.30 &    \nodata \\
 HNCO                                    &  16$_{1,16}$--15$_{1,15}$                                                                                                &  186       & 350.33306       &   0.21 $\pm$   0.02  &    4.0          & 0.89 $\pm$ 0.20      & 264.9           & 0.06 &    \nodata \\
 HNCO                                    &  16$_{2,15}$--15$_{2,14}$                                                                                                &  314       & 351.53780       &             $<$0.12  &    \nodata      &             $<$0.5   &    \nodata      & 0.06 &    \nodata \\
 HNCO                                    &  16$_{2,14}$--15$_{2,13}$                                                                                                &  314       & 351.55157       &             $<$0.12  &    \nodata      &             $<$0.5   &    \nodata      & 0.06 &    \nodata \\
 HNCO                                    &  16$_{0,16}$--15$_{0,15}$                                                                                                &  143       & 351.63326       &   0.58 $\pm$   0.03  &    4.0          & 2.50 $\pm$ 0.26      & 264.3           & 0.06 &    (12) \\
 CH$_3$CN                                &  14$_{3}$--13$_{-3}$                                                                                                     &  157       & 257.48279       &   1.18 $\pm$   0.14  &    3.7          & 4.63 $\pm$ 1.20      & 265.6           & 0.29 &    (13) \\
 CH$_3$CN                                &  14$_{2}$--13$_{2}$                                                                                                      &  121       & 257.50756       &   0.69 $\pm$   0.12  &    3.5          & 2.55 $\pm$ 0.99      & 263.9           & 0.29 &    $\dag$ (13) \\
 CH$_3$CN                                &  14$_{1}$--13$_{1}$                                                                                                      &  100       & 257.52243       &             $<$0.9  &    \nodata      &             $<$3.8   &    \nodata      & 0.29 &    (14) \\
 CH$_3$CN                                &  14$_{0}$--13$_{0}$                                                                                                      &   93       & 257.52738       &             $<$0.9  &    \nodata      &             $<$3.8   &    \nodata      & 0.29 &   \nodata \\
 CH$_3$CN                                &  19$_{4}$--18$_{4}$                                                                                                      &  282       & 349.34634       &             $<$0.12  &    \nodata      &             $<$0.5   &    \nodata      & 0.06 &    (13) \\
 CH$_3$CN                                &  19$_{3}$--18$_{-3}$                                                                                                     &  232       & 349.39330       &   0.27 $\pm$   0.02  &    3.7          & 1.08 $\pm$ 0.20      & 264.3           & 0.06 &    (14) (15) \\
 CH$_3$CN                                &  19$_{2}$--18$_{2}$                                                                                                      &  196       & 349.42685       &             $<$0.25  &    \nodata      &             $<$1.1   &    \nodata      & 0.06 &    (13) \\
 CH$_3$CN                                &  19$_{1}$--18$_{1}$                                                                                                      &  175       & 349.44699       &   0.49 $\pm$   0.02  &    2.1          & 1.11 $\pm$ 0.14      & 264.8           & 0.06 &    (13) \\
 CH$_3$CN                                &  19$_{0}$--18$_{0}$                                                                                                      &  168       & 349.45370       &   0.22 $\pm$   0.02  &    5.4          & 1.25 $\pm$ 0.28      & 264.0           & 0.06 &  \nodata \\ 
 HC$_3$N                                 &  27--26                                                                                                                  &  165       & 245.60632       &             $<$0.60  &    \nodata      &             $<$2.6   &    \nodata      & 0.30 &    \nodata \\
 SiO                                     &  6--5                                                                                                                    &   44       & 260.51801       &   0.71 $\pm$   0.11  &    6.6          & 5.02 $\pm$ 1.90      & 265.6           & 0.29 &    $\dag$ \\
 c-C$_3$H$_2$                            &  5$_{3,2}$--4$_{4,1}$                                                                                                    &   45       & 260.47975       &             $<$0.58  &    \nodata      &             $<$2.5   &    \nodata      & 0.29 &    \nodata \\
 HDO                                     &  2--1                                                                                                                    &   95       & 241.56155       &             $<$0.60  &    \nodata      &             $<$2.6   &    \nodata      & 0.30 &    \nodata \\
 C$_2$H$_5$OH                            &  6$_{5,2}$--5$_{4,1}$                                                                                                    &   49       & 340.18925       &             $<$0.12  &    \nodata      &             $<$0.5   &    \nodata      & 0.06 &    (16) \\
 C$_2$H$_5$OH                            &  20$_{2,19}$--19$_{1,18}$                                                                                                &  179       & 350.53435       &             $<$0.12  &    \nodata      &             $<$0.5   &    \nodata      & 0.06 &    \nodata \\
 C$_2$H$_5$CN                            &  16$_{5,12}$--15$_{4,11}$                                                                                                &   86       & 351.53144       &             $<$0.12  &    \nodata      &             $<$0.5   &    \nodata      & 0.06 &    \nodata \\
 CH$_3$OCH$_3$                           &  19$_{1,18}$--18$_{2,17}$ EE                                                                                             &  176       & 339.49153       &             $<$0.12  &    \nodata      &             $<$0.5   &    \nodata      & 0.06 &    (17) \\
 HCOOCH$_3$                              &  28$_{5,24}$--27$_{5,23}$ E                                                                                              &  257       & 340.74199       &             $<$0.12  &    \nodata      &             $<$0.5   &    \nodata      & 0.06 &    \nodata \\
 \textit{trans}-HCOOH                    &  15$_{3,13}$--14$_{3,12}$                                                                                                &  158       & 338.20186       &             $<$0.12  &    \nodata      &             $<$0.5   &    \nodata      & 0.06 &    \nodata \\
\enddata
\tablecomments{
Uncertainties and upper limits are of 2$\sigma$ level and do not include systematic errors due to continuum subtraction. 
Upper limits are estimated by assuming $\Delta$$V$ = 4 km s$^{-1}$. \\
$\dag$Tentative detection. 
(1) Blend with F = 4--3. 
(2) Blend with F = 3--2. 
(3) Blend of four hyperfine components. 
(4) The integrated intensity is calculated by directly integrating the spectrum.  
(5) Blend of seven hyperfine components. 
(6) Partially blended with CN at 340.03541 GHz. 
(7) Blend with F = $\frac{3}{2}$--$\frac{1}{2}$. 
(8) Blend with F = $\frac{7}{2}$--$\frac{5}{2}$ and $\frac{5}{2}$--$\frac{3}{2}$. 
(9) Blend with CH$_3$OH (4$_{0}$ E--3$_{-1}$ E). 
(10) Blend with F = $\frac{7}{2}$$^-$--$\frac{5}{2}$$^+$. 
(11) Blend with F = $\frac{5}{2}$$^+$--$\frac{3}{2}$$^-$. 
(12) Possible blend with $^{33}$SO$_2$ (5$_{4,2}$--5$_{3,3}$). 
(13) Blend of two hyperfine components. 
(14) Blend of four hyperfine components. 
(15) Partially blended with CCH at 349.39928 GHz. 
(16) Blend with 6$_{5,1}$--5$_{4,2}$. 
(17) Blend with AA, AE, and EA transitions. 
}
\end{deluxetable*}

\startlongtable
\begin{deluxetable*}{ l c c c c c c c c c }
\tablecaption{Line Parameters (CH$_3$OH) \label{tab_lines_CH3OH}}
\tablewidth{0pt}
\tabletypesize{\scriptsize} 
\tablehead{
\colhead{Molecule}   & \colhead{Transition}                               &       \colhead{$E_{u}$} &       \colhead{Frequency} &        \colhead{$T_{\mathrm{b}}$} &     \colhead{$\Delta$$V$} &     \colhead{$\int T_{\mathrm{b}} dV$} &       \colhead{$V_{\mathrm{LSR}}$} &        \colhead{RMS} &       \colhead{Note} \\
\colhead{ }          & \colhead{ }                                        &        \colhead{(K)} &           \colhead{(GHz)} &             \colhead{(K)} &          \colhead{(km s$^{-1}$)} &             \colhead{(K km s$^{-1}$)} &          \colhead{(km s$^{-1}$)} &        \colhead{(K)} &           \colhead{}
}
\startdata 
 CH$_3$OH                                &  5$_{0}$ E--4$_{0}$ E                                                                                                    &   48       & 241.70016       &   1.40 $\pm$   0.12  &    4.0          & 5.91 $\pm$ 1.27      & 263.6           & 0.30 &    \nodata \\
 CH$_3$OH                                &  5$_{-1}$ E--4$_{-1}$ E                                                                                                  &   40       & 241.76723       &   1.58 $\pm$   0.13  &    4.1          & 6.89 $\pm$ 1.25      & 264.2           & 0.30 &    \nodata \\
 CH$_3$OH                                &  5$_{0}$ A$^+$--4$_{0}$ A$^+$                                                                                            &   35       & 241.79135       &   1.44 $\pm$   0.14  &    3.7          & 5.67 $\pm$ 1.23      & 264.8           & 0.30 &    \nodata \\
 CH$_3$OH                                &  5$_{-4}$ E--4$_{-4}$ E                                                                                                  &  123       & 241.81325       &             $<$0.60  &    \nodata      &             $<$2.6   &    \nodata      & 0.30 &    \nodata \\
 CH$_3$OH                                &  5$_{4}$ E--4$_{4}$ E                                                                                                    &  131       & 241.82963       &             $<$0.60  &    \nodata      &             $<$2.6   &    \nodata      & 0.30 &    \nodata \\
 CH$_3$OH                                &  5$_{3}$ A$^+$--4$_{3}$ A$^+$                                                                                            &   85       & 241.83272       &   1.53 $\pm$   0.14  &    3.1          & 5.02 $\pm$ 1.11      & 264.4           & 0.30 &    (1) \\
 CH$_3$OH                                &  5$_{2}$ A$^-$--4$_{2}$ A$^-$                                                                                            &   73       & 241.84228       &   1.11 $\pm$   0.14  &    4.6          & 5.39 $\pm$ 1.47      & 264.3           & 0.30 &    (2) \\
 CH$_3$OH                                &  5$_{-3}$ E--4$_{-3}$ E                                                                                                  &   98       & 241.85230       &             $<$0.60  &    \nodata      &             $<$2.6   &    \nodata      & 0.30 &    \nodata \\
 CH$_3$OH                                &  5$_{1}$ E--4$_{1}$ E                                                                                                    &   56       & 241.87903       &   1.05 $\pm$   0.14  &    4.3          & 4.80 $\pm$ 1.38      & 264.8           & 0.30 &    \nodata \\
 CH$_3$OH                                &  5$_{2}$ A$^+$--4$_{2}$ A$^+$                                                                                            &   73       & 241.88767       &   0.86 $\pm$   0.15  &    1.7          & 1.60 $\pm$ 0.70      & 264.3           & 0.30 &    $\dag$ \\
 CH$_3$OH                                &  5$_{-2}$ E--4$_{-2}$ E                                                                                                  &   61       & 241.90415       &   1.48 $\pm$   0.13  &    5.8          & 9.04 $\pm$ 1.81      & 265.6           & 0.30 &    (3) \\
 CH$_3$OH                                &  14$_{-1}$ E--13$_{-2}$ E                                                                                                &  249       & 242.44608       &             $<$0.60  &    \nodata      &             $<$2.6   &    \nodata      & 0.30 &    \nodata \\
 CH$_3$OH                                &  5$_{1}$ A$^-$--4$_{1}$ A$^-$                                                                                            &   50       & 243.91579       &   1.45 $\pm$   0.12  &    4.0          & 6.10 $\pm$ 1.20      & 264.3           & 0.30 &    \nodata \\
 CH$_3$OH                                &  9$_{1}$ E--8$_{0}$ E , $\nu_t$ = 1                                                                                      &  396       & 244.33798       &             $<$0.60  &    \nodata      &             $<$2.6   &    \nodata      & 0.30 &    \nodata \\
 CH$_3$OH                                &  7$_{3}$ E--6$_{3}$ E , $\nu_t$ = 1                                                                                      &  482       & 337.51914       &             $<$0.12  &    \nodata      &             $<$0.5   &    \nodata      & 0.06 &    \nodata \\
 CH$_3$OH                                &  7$_{-2}$ E--6$_{-2}$ E , $\nu_t$ = 1                                                                                    &  429       & 337.60529       &             $<$0.12  &    \nodata      &             $<$0.5   &    \nodata      & 0.06 &    \nodata \\
 CH$_3$OH                                &  7$_{2}$ A$^+$--6$_{2}$ A$^+$ , $\nu_t$ = 1                                                                              &  363       & 337.62575       &   0.19 $\pm$   0.02  &    2.8          & 0.58 $\pm$ 0.16      & 263.2           & 0.06 &    $\dag$ \\
 CH$_3$OH                                &  7$_{2}$ A$^-$--6$_{2}$ A$^-$ , $\nu_t$ = 1                                                                              &  364       & 337.63575       &             $<$0.12  &    \nodata      &             $<$0.5   &    \nodata      & 0.06 &    \nodata \\
 CH$_3$OH                                &  7$_{0}$ E--6$_{0}$ E , $\nu_t$ = 1                                                                                      &  365       & 337.64391       &   0.27 $\pm$   0.02  &    4.6          & 1.32 $\pm$ 0.25      & 264.9           & 0.06 &    (4) \\
 CH$_3$OH                                &  7$_{3}$ A$^+$--6$_{3}$ A$^+$ , $\nu_t$ = 1                                                                              &  461       & 337.65520       &   0.18 $\pm$   0.03  &    2.3          & 0.45 $\pm$ 0.14      & 264.9           & 0.06 &    $\dag$ (5) \\
 CH$_3$OH                                &  7$_{-1}$ E--6$_{-1}$ E , $\nu_t$ = 1                                                                                    &  478       & 337.70757       &             $<$0.12  &    \nodata      &             $<$0.5   &    \nodata      & 0.06 &    \nodata \\
 CH$_3$OH                                &  7$_{0}$ A$^+$--6$_{0}$ A$^+$ , $\nu_t$ = 1                                                                              &  488       & 337.74883       &             $<$0.12  &    \nodata      &             $<$0.5   &    \nodata      & 0.06 &    \nodata \\
 CH$_3$OH                                &  7$_{1}$ A$^-$--6$_{1}$ A$^-$ , $\nu_t$ = 1                                                                              &  390       & 337.96944       &   0.24 $\pm$   0.03  &    1.5          & 0.38 $\pm$ 0.09      & 265.0           & 0.06 &    \nodata \\
 CH$_3$OH                                &  7$_{0}$ E--6$_{0}$ E                                                                                                    &   78       & 338.12449       &   1.44 $\pm$   0.03  &    3.6          & 5.54 $\pm$ 0.23      & 264.6           & 0.06 &    \nodata \\
 CH$_3$OH                                &  7$_{-1}$ E--6$_{-1}$ E                                                                                                  &   71       & 338.34459       &   1.67 $\pm$   0.03  &    4.1          & 7.25 $\pm$ 0.25      & 264.3           & 0.06 &    \nodata \\
 CH$_3$OH                                &  7$_{6}$ E--6$_{6}$ E                                                                                                    &  244       & 338.40461       &             $<$0.40  &    \nodata      &             $<$1.7   &    \nodata      & 0.06 &    \nodata \\
 CH$_3$OH                                &  7$_{0}$ A$^+$--6$_{0}$ A$^+$                                                                                            &   65       & 338.40870       &   1.73 $\pm$   0.03  &    4.8          & 8.75 $\pm$ 0.28      & 264.7           & 0.06 &    (6) \\
 CH$_3$OH                                &  7$_{-6}$ E--6$_{-6}$ E                                                                                                  &  254       & 338.43097       &             $<$0.12  &    \nodata      &             $<$0.5   &    \nodata      & 0.06 &    \nodata \\
 CH$_3$OH                                &  7$_{6}$ A$^+$--6$_{6}$ A$^+$                                                                                            &  259       & 338.44237       &   0.20 $\pm$   0.02  &    2.7          & 0.57 $\pm$ 0.14      & 265.3           & 0.06 &    (7) \\
 CH$_3$OH                                &  7$_{-5}$ E--6$_{-5}$ E                                                                                                  &  189       & 338.45654       &   0.58 $\pm$   0.03  &    2.0          & 1.25 $\pm$ 0.20      & 264.5           & 0.06 &    \nodata \\
 CH$_3$OH                                &  7$_{5}$ E--6$_{5}$ E                                                                                                    &  201       & 338.47523       &   0.37 $\pm$   0.03  &    2.9          & 1.12 $\pm$ 0.22      & 265.2           & 0.06 &    \nodata \\
 CH$_3$OH                                &  7$_{5}$ A$^+$--6$_{5}$ A$^+$                                                                                            &  203       & 338.48632       &   0.69 $\pm$   0.03  &    2.6          & 1.89 $\pm$ 0.18      & 264.7           & 0.06 &    (8) \\
 CH$_3$OH                                &  7$_{-4}$ E--6$_{-4}$ E                                                                                                  &  153       & 338.50407       &   0.77 $\pm$   0.02  &    2.7          & 2.24 $\pm$ 0.18      & 264.5           & 0.06 &    \nodata \\
 CH$_3$OH                                &  7$_{2}$ A$^-$--6$_{2}$ A$^-$                                                                                            &  103       & 338.51285       &   1.63 $\pm$   0.03  &    3.8          & 6.63 $\pm$ 0.23      & 264.6           & 0.06 &    (9) \\
 CH$_3$OH                                &  7$_{4}$ E--6$_{4}$ E                                                                                                    &  161       & 338.53026       &   0.62 $\pm$   0.03  &    2.7          & 1.75 $\pm$ 0.18      & 264.5           & 0.06 &    \nodata \\
 CH$_3$OH                                &  7$_{3}$ A$^+$--6$_{3}$ A$^+$                                                                                            &  115       & 338.54083       &   1.50 $\pm$   0.02  &    5.4          & 8.60 $\pm$ 0.30      & 263.5           & 0.06 &    (10) \\
 CH$_3$OH                                &  7$_{-3}$ E--6$_{-3}$ E                                                                                                  &  128       & 338.55996       &   0.78 $\pm$   0.03  &    3.5          & 2.89 $\pm$ 0.21      & 264.5           & 0.06 &    \nodata \\
 CH$_3$OH                                &  7$_{3}$ E--6$_{3}$ E                                                                                                    &  113       & 338.58322       &   0.90 $\pm$   0.03  &    3.3          & 3.21 $\pm$ 0.20      & 264.8           & 0.06 &    \nodata \\
 CH$_3$OH                                &  7$_{2}$ A$^+$--6$_{2}$ A$^+$                                                                                            &  103       & 338.63980       &   1.15 $\pm$   0.03  &    3.4          & 4.20 $\pm$ 0.22      & 264.5           & 0.06 &    \nodata \\
 CH$_3$OH                                &  7$_{-2}$ E--6$_{-2}$ E                                                                                                  &   91       & 338.72290       &   1.99 $\pm$   0.03  &    4.5          & 9.52 $\pm$ 0.28      & 265.0           & 0.06 &    (11) \\
 CH$_3$OH                                &  2$_{2}$ A$^+$--3$_{1}$ A$^+$                                                                                            &   45       & 340.14114       &             $<$0.12  &    \nodata      &             $<$0.5   &    \nodata      & 0.06 &    \nodata \\
 CH$_3$OH                                &  16$_{6}$ A$^-$--17$_{5}$ A$^-$                                                                                          &  509       & 340.39366       &             $<$0.12  &    \nodata      &             $<$0.5   &    \nodata      & 0.06 &    (12) \\
 CH$_3$OH                                &  11$_{1}$ E--10$_{0}$ E , $\nu_t$ = 1                                                                                    &  444       & 340.68397       &             $<$0.12  &    \nodata      &             $<$0.5   &    \nodata      & 0.06 &    \nodata \\
 CH$_3$OH                                &  4$_{0}$ E--3$_{-1}$ E                                                                                                   &   36       & 350.68766       &   1.66 $\pm$   0.03  &    3.0          & 5.26 $\pm$ 0.31      & 264.4           & 0.06 &    (13) \\
 CH$_3$OH                                &  1$_{1}$ A$^+$--0$_{0}$ A$^+$                                                                                            &   17       & 350.90510       &   1.86 $\pm$   0.02  &    2.9          & 5.68 $\pm$ 0.20      & 264.6           & 0.06 &    \nodata \\
 CH$_3$OH                                &  9$_{5}$ E--10$_{4}$ E                                                                                                   &  241       & 351.23648       &             $<$0.12  &    \nodata      &             $<$0.5   &    \nodata      & 0.06 &    \nodata \\
\enddata
\tablecomments{
Uncertainties and upper limits are of 2$\sigma$ level and do not include systematic errors due to continuum subtraction. 
Upper limits are estimated assuming $\Delta$$V$ = 4 km s$^{-1}$. \\
$\dag$Tentative detection. 
(1) Blend with 5$_{3}$ A$^-$--4$_{3}$ A$^-$. 
(2) Blend with 5$_{3}$E--4$_{3}$E. 
(3) Blend with 5$_{2}$ E--4$_{2}$ E. 
(4) Blend with 7$_{1}$ E--6$_{1}$ E. 
(5) Blend with 7$_{3}$ A$^-$--6$_{3}$ A$^-$. 
(6) Possible blend with 7$_{6}$ E--6$_{6}$ E. 
(7) Blend with 7$_{6}$ A$^-$--6$_{6}$ A$^-$. 
(8) Blend with 7$_{5}$ A$^-$--6$_{5}$ A$^-$. 
(9) Blend with 7$_{4}$ A$^-$--6$_{4}$ A$^-$ and 7$_{4}$ A$^+$--6$_{4}$ A$^+$. 
(10) Blend with 7$_{3}$ A$^-$--6$_{3}$ A$^-$. 
(11) Blend with 7$_{+2}$ E--6$_{+2}$ E. 
(12) Blend with 16$_{6}$ A$^+$--17$_{5}$ A$^+$. 
(13) Partially blended with NO at 350.68949 GHz and 350.69477 GHz. 
}
\end{deluxetable*}

\startlongtable
\begin{deluxetable*}{ l c c c c c c c c c }
\tablecaption{Line Parameters (SO$_2$, $^{34}$SO$_2$, and $^{33}$SO$_2$) \label{tab_lines_SO2}}
\tablewidth{0pt}
\tabletypesize{\scriptsize} 
\tablehead{
\colhead{Molecule}   & \colhead{Transition}                               &       \colhead{$E_{u}$} &       \colhead{Frequency} &        \colhead{$T_{\mathrm{b}}$} &     \colhead{$\Delta$$V$} &     \colhead{$\int T_{\mathrm{b}} dV$} &       \colhead{$V_{\mathrm{LSR}}$} &        \colhead{RMS} &       \colhead{Note} \\
\colhead{ }          & \colhead{ }                                        &        \colhead{(K)} &           \colhead{(GHz)} &             \colhead{(K)} &          \colhead{(km s$^{-1}$)} &             \colhead{(K km s$^{-1}$)} &          \colhead{(km s$^{-1}$)} &        \colhead{(K)} &           \colhead{}
}
\startdata 
 SO$_2$                                  &  5$_{2,4}$--4$_{1,3}$                                                                                                    &   24       & 241.61580       &   3.41 $\pm$   0.14  &    6.0          & 21.88 $\pm$ 1.98     & 264.8           & 0.30 &    \nodata \\
 SO$_2$                                  &  5$_{4,2}$--6$_{3,3}$                                                                                                    &   53       & 243.08765       &             $<$0.60  &    \nodata      &             $<$2.6   &    \nodata      & 0.30 &    \nodata \\
 SO$_2$                                  &  26$_{8,18}$--27$_{7,21}$                                                                                                &  480       & 243.24543       &             $<$0.60  &    \nodata      &             $<$2.6   &    \nodata      & 0.30 &    \nodata \\
 SO$_2$                                  &  14$_{0,14}$--13$_{1,13}$                                                                                                &   94       & 244.25422       &   4.18 $\pm$   0.14  &    5.9          & 26.26 $\pm$ 1.94     & 265.0           & 0.30 &    \nodata \\
 SO$_2$                                  &  26$_{3,23}$--25$_{4,22}$                                                                                                &  351       & 245.33923       &   1.60 $\pm$   0.14  &    5.6          & 9.54 $\pm$ 1.81      & 264.4           & 0.30 &    \nodata \\
 SO$_2$                                  &  10$_{3,7}$--10$_{2,8}$                                                                                                  &   73       & 245.56342       &   3.43 $\pm$   0.14  &    6.0          & 21.93 $\pm$ 1.96     & 265.1           & 0.30 &    \nodata \\
 SO$_2$                                  &  7$_{3,5}$--7$_{2,6}$                                                                                                    &   48       & 257.09997       &   3.59 $\pm$   0.12  &    6.7          & 25.55 $\pm$ 1.95     & 264.9           & 0.29 &    \nodata \\
 SO$_2$                                  &  32$_{4,28}$--32$_{3,29}$                                                                                                &  531       & 258.38872       &   1.23 $\pm$   0.12  &    7.2          & 9.37 $\pm$ 2.07      & 264.4           & 0.29 &    \nodata \\
 SO$_2$                                  &  20$_{7,13}$--21$_{6,16}$                                                                                                &  313       & 258.66697       &   1.08 $\pm$   0.13  &    3.7          & 4.22 $\pm$ 1.12      & 265.4           & 0.29 &    \nodata \\
 SO$_2$                                  &  9$_{3,7}$--9$_{2,8}$                                                                                                    &   63       & 258.94220       &   3.10 $\pm$   0.14  &    5.9          & 19.61 $\pm$ 1.84     & 265.0           & 0.29 &    \nodata \\
 SO$_2$                                  &  30$_{4,26}$--30$_{3,27}$                                                                                                &  472       & 259.59945       &   1.27 $\pm$   0.12  &    8.0          & 10.79 $\pm$ 2.35     & 263.9           & 0.29 &    \nodata \\
 SO$_2$                                  &  18$_{4,14}$--18$_{3,15}$                                                                                                &  197       & 338.30599       &   3.30 $\pm$   0.02  &    6.0          & 20.98 $\pm$ 0.32     & 264.9           & 0.06 &    \nodata \\
 SO$_2$                                  &  20$_{1,19}$--19$_{2,18}$                                                                                                &  199       & 338.61181       &   4.24 $\pm$   0.03  &    6.6          & 29.80 $\pm$ 0.38     & 264.3           & 0.06 &    (1) \\
 SO$_2$                                  &  39$_{6,34}$--40$_{3,37}$                                                                                                &  808       & 339.25959       &             $<$0.12  &    \nodata      &             $<$0.5   &    \nodata      & 0.06 &    \nodata \\
 SO$_2$                                  &  28$_{2,26}$--28$_{1,27}$                                                                                                &  392       & 340.31641       &   2.01 $\pm$   0.03  &    6.2          & 13.33 $\pm$ 0.35     & 264.6           & 0.06 &    \nodata \\
 SO$_2$                                  &  31$_{10,22}$--32$_{9,23}$                                                                                               &  701       & 349.22706       &   0.17 $\pm$   0.02  &    6.6          & 1.23 $\pm$ 0.34      & 263.1           & 0.04 &    (2) \\
 SO$_2$                                  &  10$_{6,4}$--11$_{5,7}$                                                                                                  &  139       & 350.86276       &   1.81 $\pm$   0.03  &    5.1          & 9.73 $\pm$ 0.32      & 264.6           & 0.06 &    \nodata \\
 SO$_2$                                  &  5$_{3,3}$--4$_{2,2}$                                                                                                    &   36       & 351.25722       &   4.03 $\pm$   0.03  &    5.9          & 25.17 $\pm$ 0.37     & 265.0           & 0.06 &    \nodata \\
 SO$_2$                                  &  14$_{4,10}$--14$_{3,11}$                                                                                                &  136       & 351.87387       &   3.40 $\pm$   0.03  &    6.2          & 22.49 $\pm$ 0.38     & 264.9           & 0.06 &    \nodata \\
 $^{34}$SO$_2$                           &  14$_{0,14}$--13$_{1,13}$                                                                                                &   94       & 244.48152       &  0.55 $\pm$   0.13  &    5.9      &   3.49 $\pm$ 1.90   &    264.3      & 0.21 &    $\dag$ (2) \\
 $^{34}$SO$_2$                           &  34$_{5,29}$--34$_{4,30}$                                                                                                &  609       & 337.87269       &             $<$0.12  &    \nodata      &             $<$0.5   &    \nodata      & 0.06 &    \nodata \\
 $^{34}$SO$_2$                           &  13$_{2,12}$--12$_{1,11}$                                                                                                &   92       & 338.32036       &   0.85 $\pm$   0.02  &    4.8          & 4.38 $\pm$ 0.41      & 265.1           & 0.06 &    \nodata \\
 $^{34}$SO$_2$                           &  14$_{4,10}$--14$_{3,11}$                                                                                                &  134       & 338.78569       &   0.78 $\pm$   0.03  &    4.0          & 3.34 $\pm$ 0.23      & 264.6           & 0.06 &    \nodata \\
 $^{34}$SO$_2$                           &  34$_{4,30}$--33$_{5,29}$                                                                                                &  593       & 350.61933       &             $<$0.12  &    \nodata      &             $<$0.5   &    \nodata      & 0.06 &    \nodata \\
 $^{34}$SO$_2$                           &  21$_{4,18}$--21$_{3,19}$                                                                                                &  250       & 352.08292       &   0.27 $\pm$   0.02  &    5.8          & 1.68 $\pm$ 0.30      & 265.0           & 0.06 &    \nodata \\
 $^{33}$SO$_2$                           &  16$_{4,12}$--16$_{3,13}$                                                                                                &  164       & 339.48232       &   0.21 $\pm$   0.02  &    4.3          & 0.99 $\pm$ 0.22      & 263.9           & 0.06 &    (3) \\
 $^{33}$SO$_2$                           &  20$_{1,19}$--19$_{2,18}$                                                                                                &  199       & 340.52818       &   0.20 $\pm$   0.12  &    6.0          & 0.73 $\pm$ 0.16      & 264.8           & 0.06 &   $\dag$ (4) (5) \\
 $^{33}$SO$_2$                           &  10$_{4,6}$--10$_{3,7}$                                                                                                  &   89       & 350.30323       &   0.22 $\pm$   0.02  &    2.9          & 0.69 $\pm$ 0.16      & 265.1           & 0.06 &    (3) \\
 $^{33}$SO$_2$                           &  13$_{4,10}$--13$_{3,11}$                                                                                                &  122       & 350.78808       &             $<$0.2  &    \nodata      &             $<$1.1   &    \nodata      & 0.06 &    (3) \\
 $^{33}$SO$_2$                           &  11$_{4,8}$--11$_{3,9}$                                                                                                  &   99       & 350.99508       &   0.29 $\pm$   0.02  &    3.3          & 1.01 $\pm$ 0.17      & 265.0           & 0.06 &    (3) \\
 $^{33}$SO$_2$                           &  8$_{4,4}$--8$_{3,5}$                                                                                                    &   72       & 351.17796       &   0.16 $\pm$   0.02  &    5.1          & 0.88 $\pm$ 0.26      & 263.9           & 0.06 &   $\dag$ (3) \\
 $^{33}$SO$_2$                           &  9$_{4,6}$--9$_{3,7}$                                                                                                    &   80       & 351.28137       &   0.27 $\pm$   0.02  &    4.0          & 1.15 $\pm$ 0.21      & 265.1           & 0.06 &    (3) \\
 $^{33}$SO$_2$                           &  7$_{4,4}$--7$_{3,5}$                                                                                                    &   64       & 351.50890       &             $<$0.12  &    \nodata      &             $<$0.5   &    \nodata      & 0.06 &    (3) \\
 $^{33}$SO$_2$                           &  6$_{4,2}$--6$_{3,3}$                                                                                                    &   58       & 351.54238       &   0.21 $\pm$   0.03  &    1.9          & 0.42 $\pm$ 0.14      & 265.2           & 0.06 &    (3) \\
 $^{33}$SO$_2$                           &  5$_{4,2}$--5$_{3,3}$                                                                                                    &   52       & 351.63513       &             $<$0.20  &    \nodata      &             $<$0.9   &    \nodata      & 0.06 &    (3) \\
 $^{33}$SO$_2$                           &  4$_{4,0}$--4$_{3,1}$                                                                                                    &   48       & 351.66122       &             $<$0.12  &    \nodata      &             $<$0.5   &    \nodata      & 0.06 &    (3) \\
 $^{33}$SO$_2$                           &  17$_{4,14}$--17$_{3,15}$                                                                                                &  179       & 351.74492       &             $<$0.20  &    \nodata      &             $<$0.9   &    \nodata      & 0.06 &    (6) \\
\enddata
\tablecomments{
Uncertainties and upper limits are of 2$\sigma$ level and do not include systematic errors due to continuum subtraction. 
Upper limits are estimated assuming $\Delta$$V$ = 4 km s$^{-1}$. \\
$\dag$Tentative detection. 
(1) Possible blend with CH$_3$OH(7$_{1}$E--6$_{1}$E). 
(2) Two channels averaged upon fitting. 
(3) Blend of ten hyperfine components. 
(4) Blend of seven hyperfine components. 
(5) The integrated intensity is calculated by directly integrating the spectrum. 
(6) Blend of four hyperfine components. 
}
\end{deluxetable*}

\section{Fitted spectra}
\begin{figure*}[tp] 
\begin{center} 
\includegraphics[width=16.2cm]{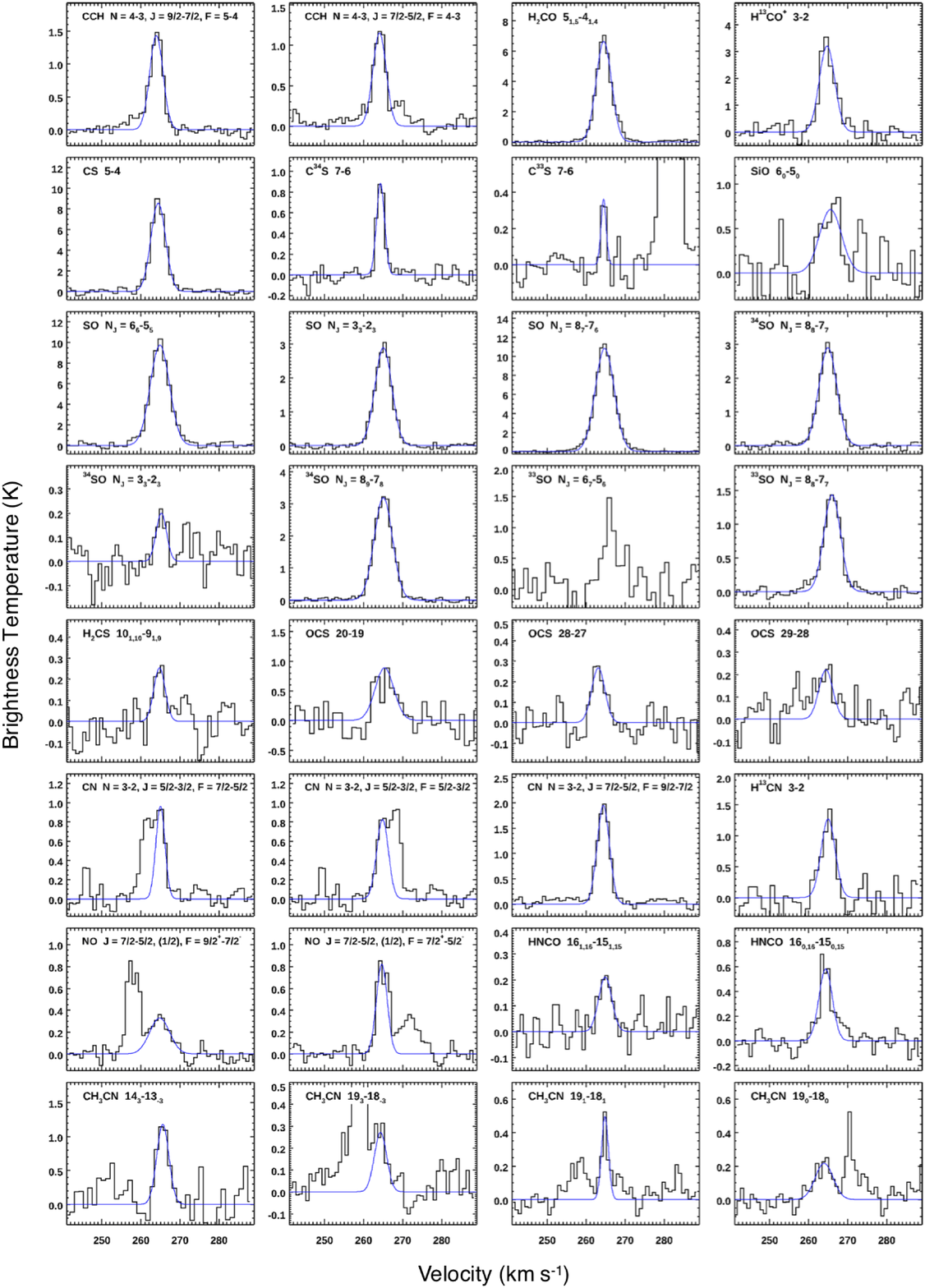} 
\caption{
ALMA spectra of CS, SO, H$^{13}$CO$^+$, OCS, $^{33}$SO, H$^{13}$CN, SiO, and CH$_3$CN lines extracted from the 0$\farcs$45 diameter region centered at ST16. 
The blue lines represent Gaussian profiles fitted to the spectra. 
For the $^{33}$SO line, the integrated intensity are derived by directly integrating the spectra between 259.0 km s$^{-1}$ and 269.5 km s$^{-1}$ (see Section \ref{sec_spc} for more details). 
}
\label{line_others}
\end{center}
\end{figure*}

\begin{figure*}[tp]
\begin{center}
\includegraphics[width=16.2cm]{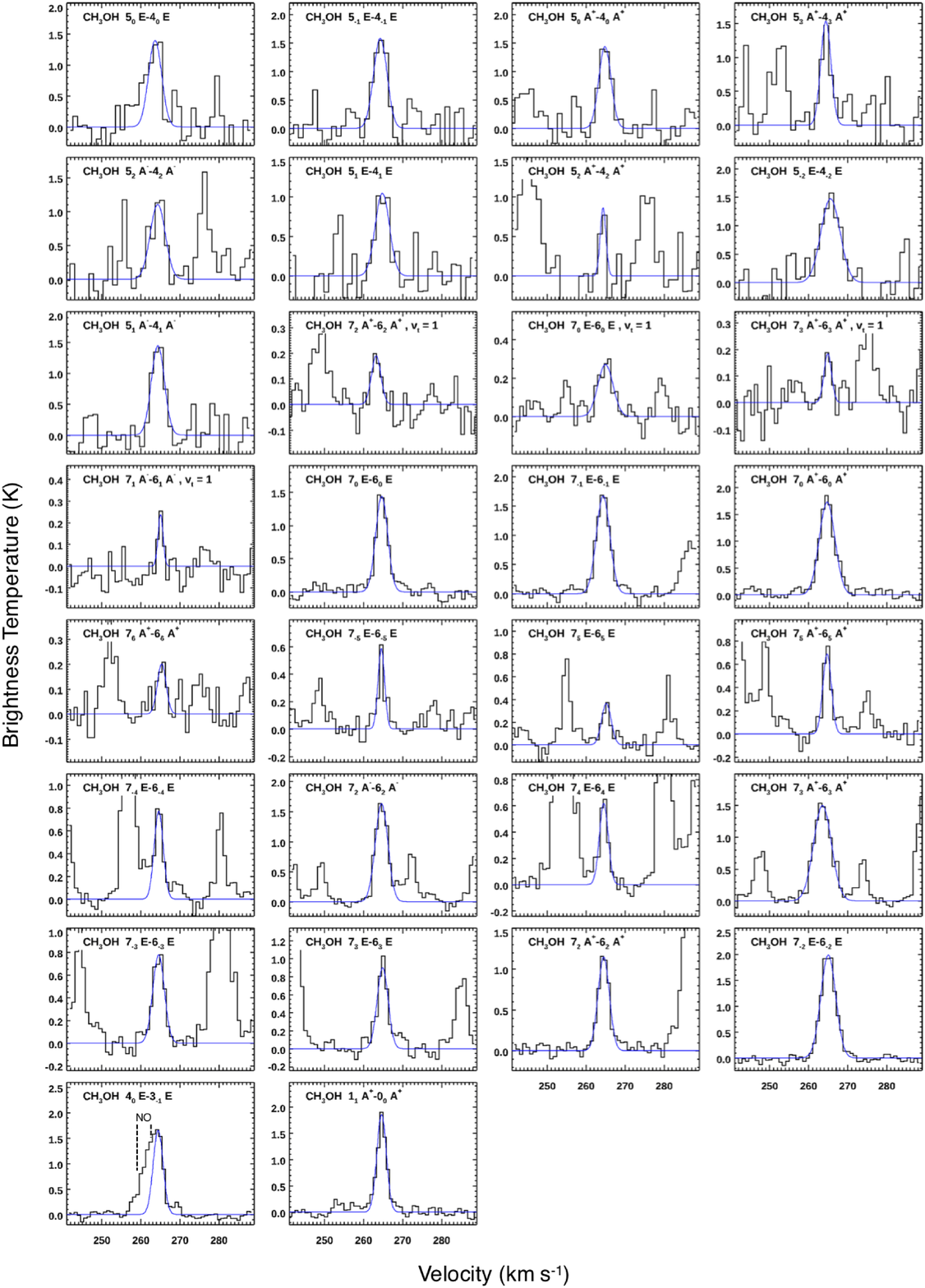}
\caption{
ALMA spectra of CH$_3$OH emission lines as in Figure \ref{line_others}. 
The spectra are sorted in ascending order of the upper state energy (the emission line with the lowest upper state energy is shown in the upper left panel and that with the highest energy is in the lower right panel). 
}
\label{line_CH3OH}
\end{center}
\end{figure*}

\begin{figure*}[tp]
\begin{center}
\includegraphics[width=16.2cm]{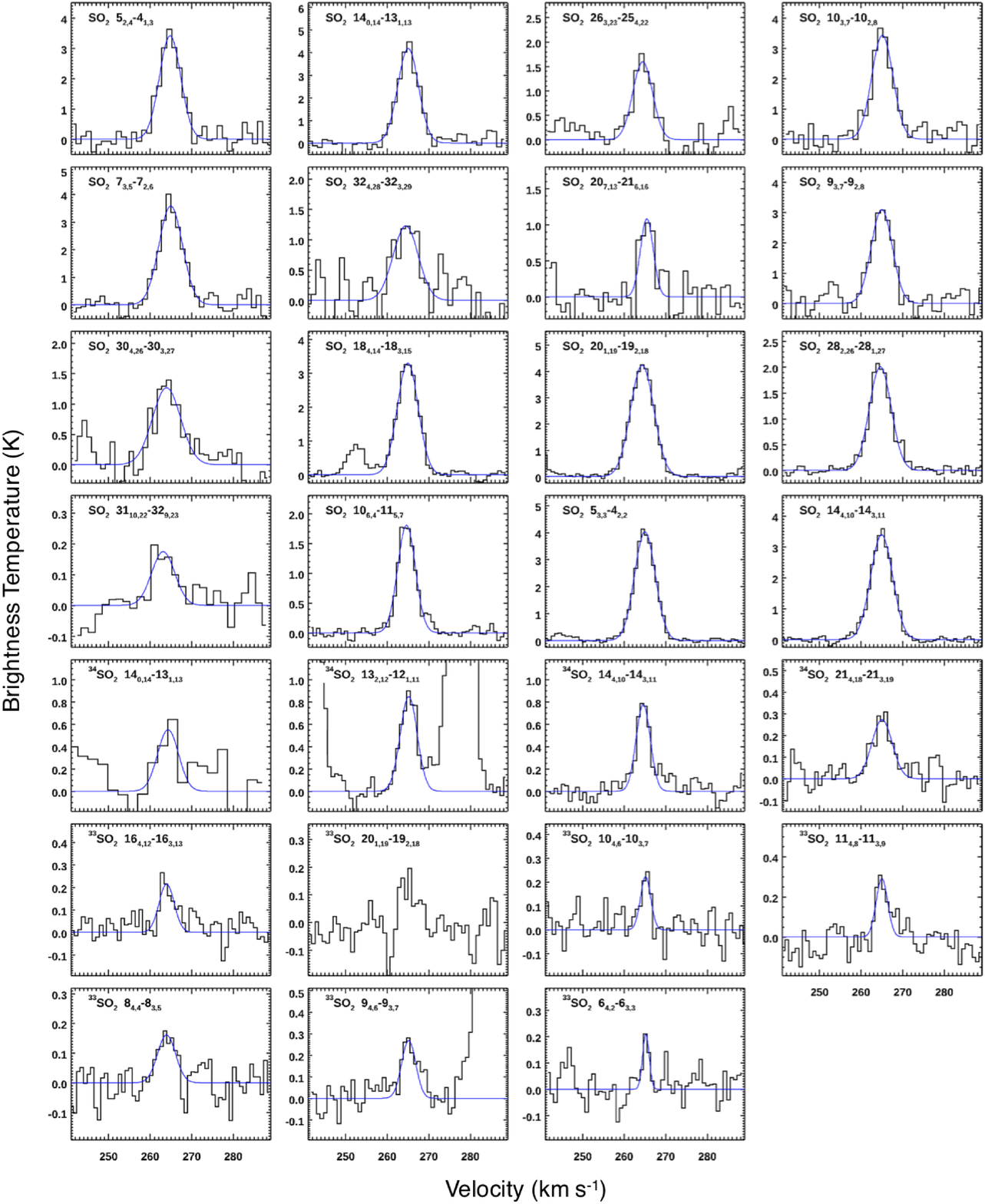}
\caption{
ALMA spectra of SO$_2$ emission lines as in Figure \ref{line_others}. 
The spectra are sorted in ascending order of the upper state energy (the emission line with the lowest upper state energy is shown in the upper left panel and that with the highest energy is in the lower right panel). 
}
\label{line_SO2}
\end{center}
\end{figure*}

\clearpage 
\restartappendixnumbering
\section{Details of the astrochemical simulations} \label{app_model} 
The present astrochemical simulations use a gas-grain chemical network code that is previously adopted in e.g. \citet{Gor17} and \citet{Sil18}. 
We have assumed that the gas and grains are coupled through accretion and various desorption mechanisms, including thermal, non-thermal \citep{Gar06}, and cosmic-ray desorption processes. 
The present gas-phase chemical network is based on the UMIST 2012 database \citep{McE13}, while the grain surface network is mainly based on \citet{Has92} and \citet{Rua16}. 

For physical conditions, we have considered three different successive evolutionary stages (i.e., cold, warm-up, and post-warm-up) as employed in e.g., \citet{Gar13}. 
The warm-up and post-warm-up stages correspond to the hot core. 
Important parameters used in the simulations are summarized in Table \ref{tab_theory1}. 
Initial elemental abundances for the LMC and Galactic cases are taken from the low-metal abundance model presented in \citet{Ach15}. 
Detailed of each evolutionary stages are described below. 

The first stage is a quiescent molecular cloud, where the cold gas chemistry and the ice mantle formation occur. 
At this stage, we consider a static molecular cloud with a gas density of $n_{\mathrm{H}}$ = 2 $\times$ 10$^4$ cm$^{-3}$. 
Grain surface chemistry is sensitive to the dust temperature and has a considerable effect on the subsequent hot core chemistry. 
Thus, visual extinction values ($A_V$) are varied from 1 mag to 5 mag in order to investigate the effect of initial physical conditions on the hot core chemistry. 
The cold stage continues up to $10^5$ years. 

Dust temperatures are related to $A_V$ and the interstellar radiation field strength using the following equation as presented in \citet{Hoc17}. 
\begin{equation} 
{T_\mathrm{d}}^{\mathrm{Hoc}}= [11 + 5.7 \tanh (0.61 - \log_{10} (A_V \cdot Z))]  {\chi_{\mathrm{uv}}}^{1/5.9},  \label{Eq_Hoc}
\end{equation} 
where $\chi_{\mathrm{uv}}$ is the Draine UV field strength \citep{Dra78}, corresponding to 1.7 $G_{\mathrm{0}}$ using the Habing field \citep{Hab68}. 
In the above equation, we have scaled down $A_V$ by the metallicity factor ($Z$), which is introduced here to mimic the metallicity effect. 
We here consider a metallicity factor of $1/3$ for the LMC simulations. 
Note that $A_V$ = 1 mag (MW) and $1/3$ mag (LMC) correspond to the gas column density of $N_{\mathrm{H_2}}$ = 2.8 $\times$ 10$^{21}$ cm$^{-2}$ using the $N_{\mathrm{H_2}}$/$A_V$ conversion factor described in Section \ref{sec_h2_sed}. 
The average interstellar radiation field strength in the LMC is suggested to be several times higher compared to the solar neighborhood based on the SED modeling of dust emission from the LMC \citep[e.g.,][]{Mei10,Gal11}. 
We here use a three times higher $\chi_{\mathrm{uv}}$ value for the LMC case.

A warm-up and collapsing stage, where the sublimation of ices occurs, follows the quiescent stage subsequently. 
At this stage, the gas density and the visual extinction gradually increases up to $n_{\mathrm{H}}$ = 6 $\times$ 10$^6$ cm$^{-3}$ and 100 mag to be consistent with the observational constraints of the ST16 hot core, as discussed in \ref{sec_h2_final}. 
The temperature gradually increases up to T$_{max}$ = 150 K, which roughly corresponds to the observed rotation temperature of the hot molecular gas (CH$_3$OH and SO$_2$, see Section \ref{sec_rd}). 
The warm-up stage continues up to $t_h = 5 \times 10^4$ years in our model. 

The final stage is a post-warm-up period, which will start just after the warm-up stage and the high-temperature chemistry proceeds. 
The post-warm-up stage continues up to 10$^5$ years, in which the density and the temperature remain the same as it was in the last phase of the warm-up stage. 
Adding up the above three stages, the total simulation time in our astrochemical model is $t_{tot} =2.5 \times 10^5$ years.

\begin{table}
\centering
{\scriptsize
\caption{Summary of the physical parameters used in the astrochemical simulations \label{tab_theory1}}
\begin{tabular}{|p{2.2in}|c|}
\hline
Grain size 	&	0.1 $\mu$m \\
Surface site density &	$1.5 \times 10^{15}$ cm$^{-2}$ \\
Gas to dust ratio	 &	100 (MW), 300 (LMC) \\
Reactive desorption factor & 0.01 \\
Ratio of the diffusion energy to the desorption energy  & 0.5 \\ 
Interstellar UV radiation field ($\chi_{\mathrm{uv}}$)	&	1 (MW), 3 (LMC) \\
Metallicity factor ($Z$) & 1 (MW), 1/3 (LMC) \\
Sticking coefficient & 0.5 \\
Cosmic ray ionization rate & $1.3 \times 10^{-17}$ s$^{-1}$ \\
Initial hydrogen number density & $2 \times 10^4$ cm$^{-3}$ \\
Final hydrogen number density & $6 \times 10^6$ cm$^{-3}$ \\
Cold stage (1$^{st}$ phase) & $10^5$ years \\
Warm-up and collapsing stage (2$^{nd}$ phase) & $5 \times 10^4$ years \\
Post-warm-up stage (3$^{rd}$ phase) & $1.0 \times 10^5$ years \\
Initial $A_V$ & $1-5$ mag \\
Final $A_V$ & 100 mag \\
${T}_{max}$ & $150$ K \\
\hline
\end{tabular}}
\end{table}

\end{document}